\documentclass[10pt,twoside,a4paper]{article}
\pdfoutput=1
\usepackage[OT2,T1,T2A]{fontenc}
\usepackage{cmap}
\usepackage{amsmath}
\usepackage{amsfonts}
\usepackage[a4paper,margin=2cm]{geometry}
\usepackage{pgf,tikz}
\usepackage{pgfplots}
\usepackage{subcaption}
\usepackage{adjustbox}
\usepackage{enumitem}
\usepackage{colortbl}
\usepackage{booktabs}
\usepackage{bm}

\usepackage[numbers]{natbib}
\makeatletter
\def\amsbb{\use@mathgroup \M@U \symAMSb}
\makeatother

\usetikzlibrary{arrows.meta}
\usetikzlibrary{decorations.markings}
\pgfplotsset{compat=newest}
\usepackage{float}
\usepackage[pdfpagelabels,plainpages=false]{hyperref}

\makeatletter
\pgfmathdeclarefunction{erf}{1}{%
	\begingroup
	\pgfmathparse{#1 > 0 ? 1 : -1}%
	\edef\sign{\pgfmathresult}%
	\pgfmathparse{abs(#1)}%
	\edef\x{\pgfmathresult}%
	\pgfmathparse{1/(1+0.3275911*\x)}%
	\edef\t{\pgfmathresult}%
	\pgfmathparse{%
		1 - (((((1.061405429*\t -1.453152027)*\t) + 1.421413741)*\t 
		-0.284496736)*\t + 0.254829592)*\t*exp(-(\x*\x))}%
	\edef\y{\pgfmathresult}%
	\pgfmathparse{(\sign)*\y}%
	\pgfmath@smuggleone\pgfmathresult%
	\endgroup
}
\makeatother
\makeatletter
\pgfmathdeclarefunction{heavi}{1}{%
	\begingroup
	\pgfmathparse{#1 > 0 ? 1 : -1}%
	\edef\sign{\pgfmathresult}%
	\pgfmathparse{abs(#1)}%
	\edef\x{\pgfmathresult}%
	\pgfmathparse{1/(1+0.3275911*\x)}%
	\edef\t{\pgfmathresult}%
	\pgfmathparse{%
		1 - (((((1.061405429*\t -1.453152027)*\t) + 1.421413741)*\t 
		-0.284496736)*\t + 0.254829592)*\t*exp(-(\x*\x))}%
	\edef\y{\pgfmathresult}%
	\pgfmathparse{(\sign)*\y}%
	\pgfmath@smuggleone\pgfmathresult%
	\endgroup
}
\makeatother

\def\cdf(#1)(#2)(#3){0.5*(1+(erf((#1-#2)/(#3*sqrt(2)))))}%
\tikzset{
	declare function={
		normcdf(\x,\m,\s)=1/(1 + exp(-0.07056*((\x-\m)/\s)^3 - 1.5976*(\x-\m)/\s));
	}
}

\pgfmathdeclarefunction{gaussfill}{3}{\pgfmathparse{#3+ 1/(#2*sqrt(2*pi))*exp(-((x-#1)^2)/(2*#2^2))}%
}

\def\hand{\includegraphics[width=3mm]{hand.png}}
\tikzset{
	-hand/.style = {
		decoration = { 
			markings, 
			mark=at position -4mm with { \node[anchor=190,transform shape, inner sep=0pt] {\hand}; } },
		postaction = decorate,
		shorten >=4mm,
	}
}

\newcommand{\aplus}{\oplus}
\graphicspath{{./Pics/}}

\title{Moate Simulation of Stochastic Processes}
\author{ Michael E. Mura\footnote{Storhaven Financial Services, United Kingdom}}
\date{31 October, 2022}

\begin{document}
\maketitle
\begin{abstract}
A novel approach called Moate Simulation is presented to provide an accurate numerical evolution of probability distribution functions represented on grids arising from stochastic differential processes where initial conditions are specified.  Where the variables of stochastic differential equations may be transformed via It\^o-Doeblin calculus into stochastic differentials with a constant diffusion term, the probability distribution function for these variables can be simulated in discrete time steps.  The drift is applied directly to a volume element of the distribution while the stochastic diffusion term is applied through the use of convolution techniques such as Fast or Discrete Fourier Transforms.  This allows for highly accurate distributions to be efficiently simulated to a given time horizon and may be employed in one, two or higher dimensional expectation integrals, e.g. for pricing of financial derivatives.  The Moate Simulation approach forms a more accurate and considerably faster alternative to Monte Carlo Simulation for many applications while retaining the opportunity to alter the distribution in mid-simulation. %end abstract
\vspace{0.5cm}


\noindent{\bf MSC-class:} 65C30 (Primary), 60G99, 91G60 (Secondary)

\noindent{\bf ACM-class:} G.3; I.6.8

\noindent{\bf JEL-class:} C15, C63, G12, G13, G17
\end{abstract}

In this paper we consider an alternative numerical approach to approximating the evolving probability density distribution for a process described by a generalised It\^o process.  This approach is called Moate Simulation to distinguish it from other approaches like Monte Carlo Simulation, Multinomial Trees and the like.   

Stochastic processes have been employed in a wide number of models and applications in finance, science and engineering.
Simply put, a stochastic process describes the indexed evolution of a random phenomenon, where in this note the index is taken to be time. The phenomenon is described as a set of random variables at each time index.  The process is often described with a set of `spatial' dimensions $X$ and a time dimension $t$. The time index can either be discrete or continuous.  The set of random  variables $\{X_t\}$ may also be  discrete or continuous.  Different cases of model can be classified:
\begin{enumerate}[label=(\Alph*)]
\item space is discrete and time is discrete;\label{discsdisct}
\item space is discrete and time is continuous;\label{discscontt}
\item space is continuous and time is discrete;\label{contsdisct}
\item space is continuous and time is also continuous.\label{contscontt}	
\end{enumerate}
A realisation of a random variable over time forms a path.  The density of all possible paths forms an evolution of  distribution of the random variables that evolves from one instant to the next, describing an evolution of probability density function over time. The initial distribution of the spatial location needs to be specified and in many cases it is known to be at a discrete point or may given by a starting distribution.   The probability distribution at a given future time is often the desired outcome for many applications, while for other applications the path dependency may be important too. 

Formal presentations of probability theory related to random processes can be found in many fine sources including \citet{doob1953stochastic},  \citet*{grimmett2001probability}, and \citet{shreve04} and so such details are not included here.  The Moate Simulation of probability density for a set of stochastic processes is general  and is applicable in many domains but, in this note, we focus on some of its financial applications.

This paper is summarised as follows.  The first section considers the generalised It\^o process in finance and discusses the path-distribution duality in existing  numerical approaches.  The next section considers a number of different ways a single discrete stochastic time step can be modelled probabilistically, and is followed by a section which discusses a set of transformations that can be applied to a probability distribution during a single step.   The details of the Moate Simulation approach are then presented in terms of these transformations.
A subsequent section describes the numerical approach to the Moate Simulation and compares an example well-known vanilla option analytic results from the Black-Scholes-Merton formulae\cite{BlackScholes1973,Merton1973} with those results generated by corresponding Moate and Monte Carlo simulations.  Finally, some extensions that are of particular interest in finance are presented for Moate Simulation, including higher dimensional modelling.

\section{Stochastic Processes for Finance}
A wide variety of stochastic processes are employed in financial modelling, for example, to describe asset prices, interest rates, volatility or forecasts of economic variables.  
When considering an appropriate model for a process, continuity or discreteness of the spatial and temporal representation needs to be established.  Processes can often start out with discrete descriptions of the variables to help elucidate the properties of the system and are later remodelled to be continuous, or vice-versa. % here we start discrete and later consider the continuous cases.

For example, in a financial setting, end-of-day prices $P$ are discrete in both space (i.e. price) and time (case \ref{discsdisct}).  If prices were purely deterministic then one might choose to represent changes in the prices with $\Delta P_t$ as
\begin{equation}
	\Delta P_t = \mu(P_t,t) \Delta t \label{eqn:discretetime}
\end{equation} 
where $\mu(P_t,t)$ is some function of space and time.
The usual notation of calculus is employed here to represent a discrete change in $a$ as $\Delta a$ and continuous change in $a$ to be represented as an infinitesimal change $da$. Also, the index $t$ is represented as a subscript to the spatial variable, here $P_t$.

Discrete prices, despite representing reality, are rarely modelled as such so the spatial dimension is often approximated to be continuous, case \ref{contsdisct} in the above classification.  The same equation (\ref{eqn:discretetime})  can be used for continuous prices with a discrete time step.  The case \ref{contscontt}  is often used when prices need to be modelled intra-day where the continuous time is convenient when modelling.  The notation for this is the familiar expression for a deterministic continuous (space and time) process 
\begin{equation}
	dP_t = \mu(P_t,t) \; dt. \label{eqn:continuoustime}
\end{equation}

Deterministic processes in finance are not as common as stochastic (ergo non-deterministic) processes.  
The generalised  It\^{o} process \cite{Ito1944} is given by
\begin{equation}
	dP_t = a(P_t,t) \;dt  + b(P_t,t)\; dW_t.\label{eqn:generalito}
\end{equation}
where the terms $a$ and $b$ are both functions of the stochastic variable $P_t$ at time $t$  and time $t$ itself.  
\citet{Bachelier1900} produced one of the earliest examples of continuous time analysis, introducing the idea that stocks had a random nature that could be modelled with Brownian Motions. \citet{Samuelson1965} introduced the stochastic differential equation for log-normal processes in the form familiar to  practitioners of financial economics
\begin{equation}
	dP_t = P_t \mu \; dt + P_t \sigma \; dW_t \label{eqn:samuelson}
\end{equation}    
This stochastic differential equation is an axiom of the Black-Scholes-Merton model\cite{BlackScholes1973,Merton1973}.  The Samuelson stochastic differential equation \ref{eqn:samuelson} contains two constants, $\mu$ and $\sigma$ and two differentials terms, $dt$ and the Wiener increment $dW_t$, thus $dP_t$ is an It\^o process.  The first term in the equation represents a (deterministic) drift term and the second term a stochastic diffusion term.   The Wiener increment $dW_t$ can be considered to be a Gaussian function with infinitesimal width, in effect a normal distribution ${\cal{N}}(0,dt)$ with a variance of $dt$.  This decomposition of the It\^{o} process into a deterministic drift term and a stochastic diffusion term has influenced the way that financial practitioners think about price moves in financial instruments.  The Samuelson stochastic differential equation \ref{eqn:samuelson} has formed a standard model for equity processes, while a large variety of other It\^{o} processes exist  for interest rates (Vasicek, Ornstein-Uhlenbeck, et al.)  and stochastic volatility (for examples see \citet{BrigoMercurio2001}). 

A kind of duality exists in the description of the evolution of stochastic variables; they may be considered as paths, or as probability distributions which evolve over time.
For financial practitioners it is very natural to think of the price processes in terms of paths because historical prices are typically presented as time series.  This means that the  Monte Carlo approach introduced to finance by Boyle \cite{Boyle1977} is so intuitive that stochastic processes are very often and predominately imagined as a set of Monte Carlo paths for modelling purposes.  Binomial (or multinomial) trees are another useful alternative for modelling stochastic processes (see \citet{shreve2004stochastic1})  where each set of the stochastic variables at a time $t$ are given as a discrete set of spatial values with associated probabilities. The tree approaches identify a set of specific paths and then determine the probability of being at any node on the tree at any given time.   

The evolving probability distribution view is more natural where a partial differential equation (PDE) can be derived from the stochastic differential  equation.  Using suitable boundary conditions, analytic or numerical solutions of the PDE may sometimes be obtained which determine the probability distribution of the stochastic variable either explicitly or implicitly.  In general, analytic solutions may be difficult to obtain, and the numerical solutions can be computationally expensive where they are tractable.

In many cases the goal of the modelling is to obtain the probability distribution of the stochastic variable at a particular time and employ it in an expectation integral.  The different modelling approaches to stochastic processes have their own strengths and weaknesses, typically in accuracy, flexibility,  expense of computation or accepted model risk.

With respect to the path-distribution duality, the Moate Simulation approach is closest to the probability distribution view.  The approach is to start with a known value or distribution describing the stochastic variable and evolve the distribution with discrete time steps according to the dynamics of the given stochastic differential equation; the discrete time aspect is similar to Monte Carlo or Tree approaches. With Moate Simulation the distribution is altered between two adjacent time steps by moving the distribution with the drift term and then applying diffusion to drifted distribution (or alternatively diffusing first then drifting).  The probability distribution is represented as a function on a set of grid points and the accuracy depends on the fineness of the grid used.   

Next a number of different ways in which a single step is simulated is presented as found in some of the models already discussed.

\pgfmathdeclarefunction{gauss}{3}{%
	\pgfmathparse{1/(#3*sqrt(2*pi))*exp(-((#1-#2)^2)/(2*#3^2))}%
}

\newcommand\tzero{0}
\newcommand\tone{0.7}
\newcommand\thalf{0.35}
\begin{figure}[!t]
\begin{center}	
	
\begin{subfigure}[b]{0.3\textwidth}		
	\begin{tikzpicture}[scale=2.5] 
		\draw[dashed] (0,1) -- (0,0);  \node [below] at (0,0) {$t_0$};
		\draw[dashed] (\tone,1) -- (\tone,0);  \node [below] at (\tone,0) {$t_1$};
		\node [left] at (0,0.5) {X};  \draw[black,fill=black] (0,0.5) circle (0.1ex);
	    \draw[-{Latex[length=3mm, width=2mm]}]  (0,0.5) -- (\tone,0.75) node[midway,above] {$p=1$}; 
	      \node [right] at (\tone,0.75) {$X+\Delta X$}; \draw[black,fill=black] (\tone,0.75) circle (0.1ex);
	    \node [below] at (\thalf,-0.2) {(a) Deterministic};
	\end{tikzpicture}
	\end{subfigure}%
	\hfill
	\begin{subfigure}[b]{0.3\textwidth}	
	\begin{tikzpicture}[scale=2.5] 
	\draw[dashed] (0,1) -- (0,0);  \node [below] at (0,0) {$t_0$};
	\draw[dashed] (\tone,1) -- (\tone,0);  \node [below] at (\tone,0) {$t_1$};
	\node [left] at (0,0.5) {X};  \draw[black,fill=black] (0,0.5) circle (0.1ex);
	\draw[-{Latex[length=3mm, width=2mm]}]  (0,0.5) -- (\tone,0.75) node[midway,above] {$p_u$}; 
	  \node [right] at (\tone,0.75) {$X+\Delta X_u$}; \draw[black,fill=black] (\tone,0.75) circle (0.1ex);
	\draw[-{Latex[length=3mm, width=2mm]}]  (0,0.5) -- (\tone,0.25) node[midway,below] {$p_d$}; 
	  \node [right] at (\tone,0.25) {$X+\Delta X_d$}; \draw[black,fill=black] (\tone,0.25) circle (0.1ex);
	\node [below] at (\thalf,-0.2) {(b) Binomial};
	\end{tikzpicture}
	\end{subfigure}%
\hfill
\begin{subfigure}[b]{0.3\textwidth}	
	\begin{tikzpicture}[scale=2.5] 
	\draw[dashed] (0,1) -- (0,0);  \node [below] at (0,0) {$t_0$};
	\draw[dashed] (\tone,1) -- (\tone,0);  \node [below] at (\tone,0) {$t_1$};
	\node [left] at (0,0.5) {X};  \draw[black,fill=black] (0,0.5) circle (0.1ex);
	\draw[-{Latex[length=3mm, width=2mm]}]  (0,0.5) -- (\tone,0.8) node[midway,above] {$p_u$}; 
	  \node [right] at (\tone,0.8) {$X+\Delta X_u$}; \draw[black,fill=black] (\tone,0.8) circle (0.1ex);
	\draw[-{Latex[length=3mm, width=2mm]}]  (0,0.5) -- (\tone,0.55) node[midway,below] {$p_m$}; 
	  \node [right] at (\tone,0.55) {$X+\Delta X_u$}; \draw[black,fill=black] (\tone,0.55) circle (0.1ex);
	\draw[-{Latex[length=3mm, width=2mm]}]  (0,0.5) -- (\tone,0.2) node[midway,below] {$p_d$}; 
	  \node [right] at (\tone,0.2) {$X+\Delta X_d$}; \draw[black,fill=black] (\tone,0.2) circle (0.1ex);
    \node [below] at (\thalf,-0.2) {(c) Trinomial};
	\end{tikzpicture}
	\end{subfigure}\\ %
 \vspace{1cm}
\begin{subfigure}[b]{0.3\textwidth}	
	\begin{tikzpicture}[scale=2] 
	\draw[dashed] (0,1) -- (0,0);  \node [below] at (0,0) {$t_0$};
	\draw[dashed] (\tone,1) -- (\tone,0);  \node [below] at (\tone,0) {$t_1$};
	\node [left] at (0,0.5) {X};  \draw[black,fill=black] (0,0.5) circle (0.1ex);
	\draw[-{Latex[length=3mm, width=2mm]}]  (0,0.5) -- (\tone,0.9) node[midway,above] {$p_1$}; 
	\node [right] at (\tone,0.9) {$X+\Delta X_1$}; \draw[black,fill=black] (\tone,0.9) circle (0.1ex);	
	\foreach \i in {2,...,8}
	{
		\draw[-{Latex[length=3mm, width=2mm]}]  (0,0.5) -- (\tone,\i / 10); 
	    \draw[black,fill=black] (\tone,\i / 10) circle (0.1ex);
	}
	\node [right] at (\tone +0.2,0.55) {$\vdots$}; 
	\draw[-{Latex[length=3mm, width=2mm]}]  (0,0.5) -- (\tone,0.1) node[midway,below] {$p_n$}; 
	\node [right] at (\tone,0.1) {$X+\Delta X_n$}; \draw[black,fill=black] (\tone,0.1) circle (0.1ex);
	\node [below, label={[align=right](d) Multinomial\\ $n$}] at (\thalf+0.1,-0.8) { };
    \end{tikzpicture}	
\end{subfigure}%
\hfill
\begin{subfigure}[b]{0.3\textwidth}	
	\begin{tikzpicture}[scale=2] 
	\draw[dashed] (0,1) -- (0,0);  \node [below] at (0,0) {$t_0$};
	\draw[dashed] (\tone,1) -- (\tone,0);  \node [below] at (\tone,0) {$t_1$};
	\node [left] at (0,0.5) {X};  \draw[black,fill=black] (0,0.5) circle (0.1ex);
	\draw[-{Latex[length=2mm, width=2mm]}]  (0,0.5) -- (\tone,0.745) node[midway,above] {$p_1$}; 
	\node [right] at (\tone,0.745) {$X+\Delta X_1$}; \draw[black,fill=black] (\tone,0.745) circle (0.1ex);	
	\foreach \i in { 100, 352, 427, 501, 537, 658, 671, 712}  % had 45 at bottom and 745 at top
	{
		\draw[-{Latex[length=2mm, width=2mm]}]  (0,0.5) -- (\tone,\i / 1000); 
		\draw[black,fill=black] (\tone,\i / 1000) circle (0.1ex);
	}
	\node [right] at (\tone +0.2,0.5) {$\vdots$}; 
	\draw[-{Latex[length=2mm, width=2mm]}]  (0,0.5) -- (\tone,0.045) node[midway,below] {$p_n$}; 
	\node [right] at (\tone,0.045) {$X+\Delta X_n$}; \draw[black,fill=black] (\tone,0.045) circle (0.1ex);
	\node [below, label={[align=right](e)  Monte Carlo\\ $n=10$}] at (\thalf+0.1,-0.8) { };
    \end{tikzpicture}
	\end{subfigure}%
\hfill
\begin{subfigure}[b]{0.3\textwidth}	
	\begin{tikzpicture}[scale=2] 
	\draw[dashed] (0,1) -- (0,0);  \node [below] at (0,0) {$t_0$};
	\draw[dashed] (\tone,1) -- (\tone,0);  \node [below] at (\tone,0) {$t_1$};
	\node [left] at (0,0.5) {X};  \draw[black,fill=black] (0,0.5) circle (0.1ex);
	\draw[-{Latex[length=2mm, width=2mm]}]  (0,0.5) -- (\tone,0.885) node[midway,above] {$p_1$}; 
	\node [right] at (\tone,0.885) {$X+\Delta X_1$}; \draw[black,fill=black] (\tone,0.885) circle (0.1ex);	
	\foreach \i in {  100, 131, 149, 170, 204, 249, 254, 314, 357, 364, 369, 387, 389, 390, 449, 460, 461, 464, 473, 477, 510, 518, 525, 530, 531, 535, 536, 539, 539, 577, 582, 602, 602, 612, 626, 642, 642, 645, 650, 662, 682, 690, 691, 708, 709, 730, 777, 863}  % had 51 at bottom and 885 at top
	{
		\draw[-{Latex[length=2mm, width=2mm]}]  (0,0.5) -- (\tone,\i / 1000); 
		\draw[black,fill=black] (\tone,\i / 1000) circle (0.1ex);
	}
	\node [right] at (\tone +0.2,0.5) {$\vdots$}; 
	\draw[-{Latex[length=2mm, width=2mm]}]  (0,0.5) -- (\tone,0.051) node[midway,below] {$p_n$}; 
	\node [right] at (\tone,0.051) {$X+\Delta X_n$}; \draw[black,fill=black] (\tone,0.051) circle (0.1ex);
	\node [below, label={[align=right](f)  Monte Carlo\\ $n=50$}] at (\thalf+0.1,-0.8) { };
     \end{tikzpicture}
	\end{subfigure}\\ %
 \vspace{1cm}
\begin{subfigure}[b]{0.3\textwidth}	      
	\begin{tikzpicture}[scale=2]
	\draw[dashed] (0,1) -- (0,0);  \node [below] at (0,0) {$t_0$};
	\draw[dashed] (\tone,1) -- (\tone,0);  \node [below] at (\tone,0) {$t_1$};
	\node [left] at (0,0.5) {X};  \draw[black,fill=black] (0,0.5) circle (0.1ex);
	(2,1) coordinate (A);
	\foreach \i in {4,...,16}
	{
		\draw[gray, solid]  (0,0.5) -- (\tone,\i / 20); 
	}
	\draw[solid] (\tone+0.1,1) -- (\tone+0.1,0.8) -- (\tone+0.5,0.8)  -- (\tone+0.5,0.2) -- (\tone+0.1,0.2) -- (\tone+0.1,0);
		\node [below, label={[align=right](g) Continuous\\ Uniform}] at (\thalf+0.1,-0.8) { };
    \end{tikzpicture}
	\end{subfigure}%
\hfill
\begin{subfigure}[b]{0.3\textwidth}	
	\begin{tikzpicture}[scale=2]
		\draw[dashed] (0,1) -- (0,0);  \node [below] at (0,0) {$t_0$};
		\draw[dashed] (\tone,1) -- (\tone,0);  \node [below] at (\tone,0) {$t_1$};
		\node [left] at (0,0.5) {X};  \draw[black,fill=black] (0,0.5) circle (0.1ex);
        (2,1) coordinate (A);
        \foreach \i in {4,...,16}
        {
        	\draw[gray, solid]  (0,0.5) -- (\tone,\i / 20); 
        }
        \draw[solid] (\tone+0.1,1) -- (\tone+0.1,0.8) -- (\tone+0.5,0.5)  -- (\tone+0.1,0.2) -- (\tone+0.1,0);
		\node [below, label={[align=right](h) Continuous\\ Triangular}] at (\thalf+0.1,-0.8) { };
	\end{tikzpicture}
	\end{subfigure}%
\hfill
\begin{subfigure}[b]{0.3\textwidth}	
	\begin{tikzpicture}[scale=2]
	\draw[dashed] (0,1) -- (0,0);  \node [below] at (0,0) {$t_0$};
	\draw[dashed] (\tone,1) -- (\tone,0);  \node [below] at (\tone,0) {$t_1$};
	\node [left] at (0,0.5) {X};  \draw[black,fill=black] (0,0.5) circle (0.1ex);
	(2,1) coordinate (A);
	\foreach \i in {2,...,18}
	{
		\draw[gray, solid]  (0,0.5) -- (\tone,\i / 20); 
	}
	% Normal Distribution 2
	\begin{axis}[
		name=axis1,
		clip=false,
		samples=100,
		ymin=0,
		xmin=-0.5,xmax=2.5,
		scale only axis,
		domain=-2:2,
		rotate=270,
		hide axis,
		shift={(0.55 cm, 0.75 cm)},
		scale=0.8,
		width=2cm,
		height=1cm]
		
		\addplot[restrict x to domain=-0.45:1.45] {gauss(x, 0.5, 0.25)}	coordinate [pos=0] (g1)	coordinate [pos=1] (g2)	coordinate [pos=0.5] (gm);
	\end{axis}
		\node [below, label={[align=right](i) Continuous\\ Normal}] at (\thalf+0.1,-0.8) { };
    \end{tikzpicture}
	\end{subfigure}%
\caption{The sub-figures show examples of discrete (a-d), Monte Carlo (e-f) and continuous (g-i) outcomes for one period between $t_0$ and $t_1$ for a starting quantity $X$.  The discrete probabilities $p_i$ must sum to 1 for (a-f) and are expected to be greater than 0, while the continuous probability distribution functions integrate to unity. For the discrete cases the changes in $X$, denoted $\Delta X_i$ are distinct and finite.  As the Monte Carlo values are drawn from iid (independent, identically distributed) distributions each outcome is considered to be have the same probability $p=1/n$.\label{fig:singlesteps}}
    \end{center}
    \end{figure}
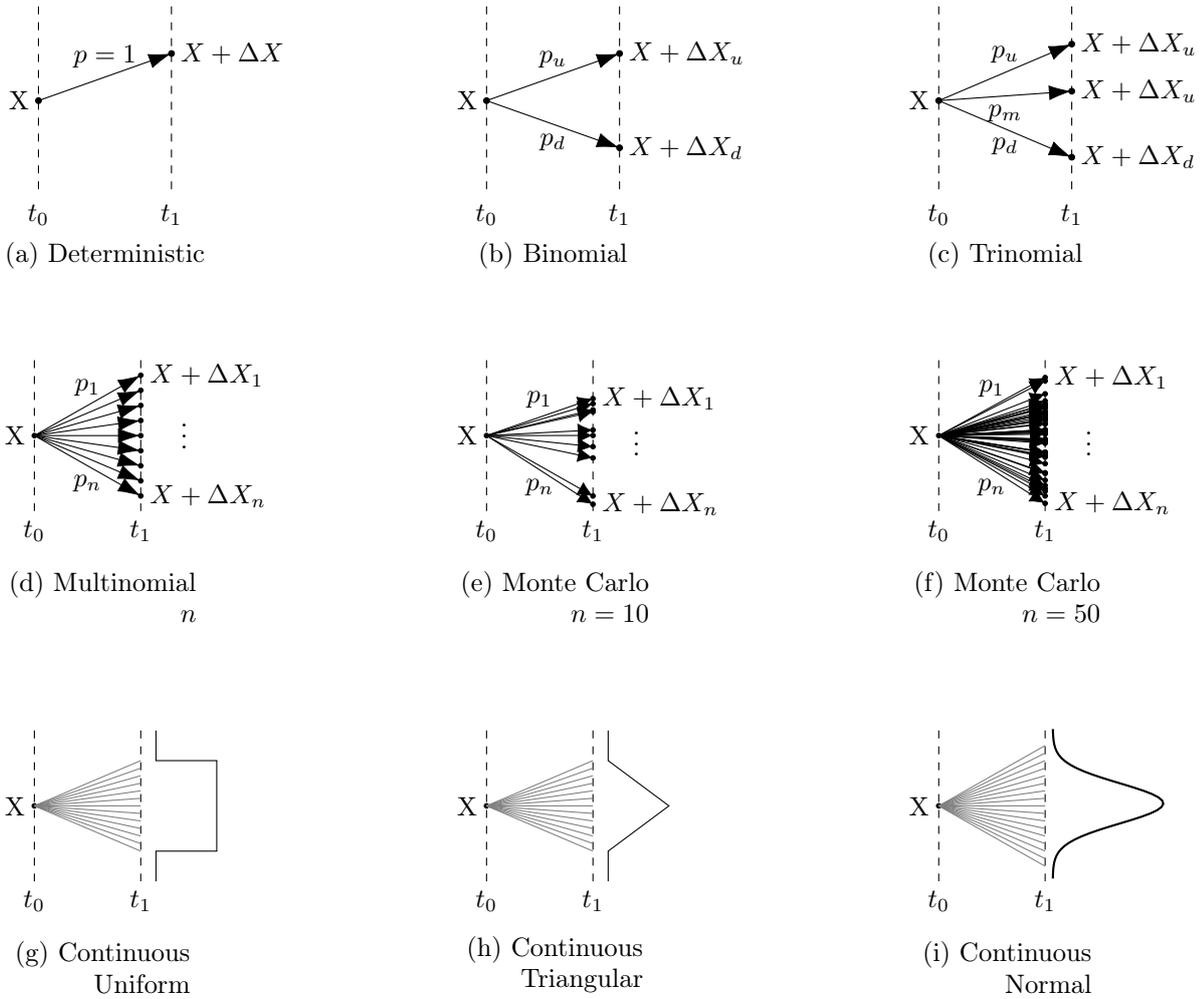

\section{Modelling a Single Time Step}

A variety of ways to represent a single time step in a continuous space, discrete time process, case \ref{contsdisct} are presented in figure \ref{fig:singlesteps}.  Each sub-figure shows a different modelling approach to a time step between time $t_0$ and time $t_1$.  At $t_1$ the values of the stochastic variable $X$ are shown along with their associated probabilities (discrete) or probability distributions (continuous).

Figure \ref{fig:singlesteps}(a) represents the simplest representation where a change occurs between the two times that is completely determined with probability $p=1$.  This is the first in the family sequence of models which includes the binomial, fig \ref{fig:singlesteps}(b), and trinomial, fig \ref{fig:singlesteps}(c), models, where a probability is assigned to each of the two or three respective outcomes, but since the change shown in \ref{fig:singlesteps}(a) is a single outcome the result is certain.  In this sense, a deterministic change is a special edge case of the multinomial description in \ref{fig:singlesteps}(d).  The use of the binomial models in derivative pricing is discussed extensively by \citet{shreve2004stochastic1}, where single steps are combined into trees to describe the stochastic processes.

Next we consider Monte Carlo approaches as introduced to finance by \citet{Boyle1977}.  Figure \ref{fig:singlesteps}(e) shows 10 different draws from an i.i.d.\ distribution, each representing a single step between $t_0$ and time $t_1$.  It is more usual to see a Monte Carlo simulation as a set of consecutive steps forming a path between two points in time and space formed by random spatial moves on a regular time grid, with this evolution being performed many times to create a set of paths.  Here we focus on a single step so that each draw creates its own spatial move over the same single time step.  Although each step is discrete, as the set of $n$ spatial moves increases the histogram of positions increasingly tends to resemble the distribution from which the random moves were drawn.  Figure \ref{fig:singlesteps}(f) shows 50 such single steps. The probability attributed to each move is typically treated as the reciprocal of the number of draws, $n$, and where the process is i.i.d.\ the increments may be weakly or strongly stationary.

Finally we consider the continuous distribution generated during a single step.
In the limit, it is possible to describe the outcome of a single time step as being represented by at least a piecewise continuous probability distribution function.   Figure \ref{fig:singlesteps}(g)  represents a uniform probability distribution between a pair of spatial limits sometimes denoted as $\Pi(X)$, figure  \ref{fig:singlesteps}(h)  represents a triangular probability distribution function sometimes denoted by $\Lambda(X)$, while figure  \ref{fig:singlesteps}(i) represents a continuous normal probability distribution function sometimes represented as ${\cal N}(\mu,\sigma^2)$ where $\mu$ is the mean and $\sigma^2$ is the variance. There is a deep connection between these three distributions.  The triangular function is formed by adding two random variables drawn from the uniform distribution (think of the outcome of rolling two dice).  In the limit, the addition of many random variables tends towards a normal probability distribution as outlined by the Central Limit Theorem. See section 3.6 of \citet{osgood2019lectures} for a more detailed discussion.  

In the class of continuous distributions, the discrete deterministic single step of figure \ref{fig:singlesteps}(a) can be considered to be a Dirac delta function, $\delta$, a continuous distribution function with infinitesimal width.

The  continuous normal probability distribution function in figure  \ref{fig:singlesteps}(i) is a discrete time representation of a Wiener increment $\Delta W$. This invites the a single step in the generalised It\^o process in equation \ref{eqn:generalito} to be seen as a drift from an origination point coupled with a diffusion yielding a distribution at the end of the time step.  This distributional view complements the Monte Carlo style path view, and it is this distributional approach that is employed in Moate Simulation. 

\section{Transforming Distributions}
A stochastic increment over a time $\Delta t$ can be characterised by one or more transformations $Z$ that turn a distribution $f_t(X)$ at $t$ into a different distribution $f_{t+\Delta t}(X)$ at $t+\Delta t$, in either the discrete or continuous context
\begin{equation}
      f_{t+\Delta t}(X) = Z f_t(X).  
\end{equation}
The distribution is equivalently described by either its probability density function (PDF) or the cumulative distribution function (CDF).  We will use $f_t(X)$ denote the CDF with the corresponding PDF is its derivative denoted here as $F_t(X) = \frac{df_t(X)}{dX}$. 

Some additional notation concerning the representation of the distribution functions on a grid is introduced here.  The function $f_t(X)$  is the distribution at time $t$ is defined on a domain of $X$. A grid representation of $f_t(X)$ is defined on a set of $n$ values of $X$ giving rise to a set of $n$ tuples of position $X_i$ and the value $f_i^{(t)} \equiv f(X_i,t)$, that is $(X^{(t)}_i,f^{(t)}_i) \forall i \in n$.  The superscript $t$ for $X$ is necessary because the grid may change from one time step to another.

There are four types of transformation of interest in this paper.
\begin{itemize}
	\item {\bf change of variable}: 
	$(X^{(t)}_i,f^{(t)}_i) \;\;\underrightarrow{\hbox{change of variable\ }} \;\;(Y^{(t)}_i,f^{(t)}_i)$;
	\item {\bf interpolation}:  $\{(X^{(t)}_i,f^{(t)}_i) \}\;\;\underrightarrow{\hbox{interpolation\ }} \;\;\{(X^{(t)}_j,f^{(t)}_j\})$;
	\item {\bf deterministic transformation}:  $(X^{(t)}_i,f^{(t)}_i) \;\;\underrightarrow{\hbox{drift\ }} \;\;(X^{(t)}_i + a(X^{(t)}_i,t) \Delta t, f^{(t)}_i)$;
	\item {\bf diffusion transformation}: 
	         $\{(X^{(t)}_i,f^{(t)}_i)\} \;\;\underrightarrow{\hbox{diffuse\ }} \;\;\{(X^{(t+1)}_i,f^{(t+1)}_i)\}$.
\end{itemize}
We consider each transformation in turn.

\subsection{Change of Variable}
Consider the change of variable from $X$ to $Y$  for the function $Y(X)$  with a Jacobian $\frac{dX}{dY}$ 
\begin{equation}
	f(X)|_r^s =\int_r^s F(X) dX = \int_{Y(r)}^{Y(s)} F(X(Y)) \frac{dX}{dY} \; dY
\end{equation}
The value of cumulative distribution remains unchanged with the change in variable so converting to the PDF takes account of the Jacobian implicitly.  

\subsection{Interpolation}
Interpolation is a transformation intended to leave the underlying function unaltered as the grid point representation is changed from one set of points, $i\in n$, to another, $j \in m$. In many respects interpolation is like an identity but it can have consequences that impact the numerical accuracy.

\subsection{Deterministic Transformation}
The deterministic transformation moves the mass of the distribution from one point $X_i$ to a different single point.  Where the drift is a constant, for example $\mu \Delta t$, then the probability mass associated with $X_i$ moves to $X_i + \mu \Delta t$, effecting a simple translation of the cumulative distribution function $f$.   Where the drift is a function of time or position, for example $\mu X^2 \Delta t$, then the probability mass moves from $X_i$ to $X_i + \mu X_i^2 \Delta t$; in this instance the CDF does not move uniformly but is stretched differently depending on the value of $X_i$.  If the general drift coefficient is $a(X,t)$ over a time step $\Delta t$ then we defined the operator $\aplus$ to take a distribution $f$ and drift of stretch the probability mass by $a(X,t)\Delta t$ which is written as $f^{(t)} \aplus a(X,t)\Delta t$;  this operator was selected by analogy to the $+$ addition operator in $X^{(t+\Delta t)} = X^{(t)} + a(X,t) \Delta t$.

\subsection{Diffusion Transformation}
The diffusion transformation above redistributes all the probability mass from each location according to some stochastic function with an associated distribution.  For a Wiener process the stochastic term is normal. 

To visualise this, consider the following analogy.  The waves from the sea propagate forward until the come on some closely spaced posts (about a wavelength in spacing).  Each wave passes through and diffracts, that is it spreads out from each gap it just passed came through in roughly semi-circular form.  The semi-circular wavefronts recombine so that at some distance from the posts the wave appears linear again.  The analogy is that, at the gap (the locus), the water in the wave (initial distribution) coming though diffracts (diffuses) in semi-circles (normal distributions) and the new wave front (final distribution).

At every moment each piece of probability mass in each contiguous locus is diffusing (and in the general case it may be drifting too).
This redistribution transformation is achieved through the convolution of the starting distribution with the diffusive distribution (for example, a normal distribution to represent a Wiener process over discrete time step), yielding the new distribution at after the step time $\Delta t$. 

It is useful to recall some of the properties of the convolution operator. 
The convolution of two distributions $f(X)$ and $g(X)$ is denoted by $\ast$, is a commutative operation and is defined as 
\begin{eqnarray}
	h(X) & = & \int_{-\infty}^{\infty} f(u) g(X -u) \;du\\
	& = & \int_{-\infty}^{\infty} f(X-u) g(u) \;du \\
	& = & f(X) \ast g(X)  = g(X) \ast f(X).
\end{eqnarray}
The addition of two independent random variables produces a distribution that is formed by convolution.  If $\Pi(X)$ represents a uniform distribution function and then convolution between two $\Pi(X)$ functions yields a triangular function $\Lambda(X)$ 
\begin{equation}
	\Lambda(X) = \Pi(X) \ast \Pi(X)
\end{equation}
The convolution of a discrete distribution characterised as a Dirac delta function $\delta_0(X)$ with a distribution $f(X)$ yields the same function $f(X)$
\begin{equation}
	f(X) = f(X) \ast \delta_0(X) = \delta_0(X) \ast f(X)
\end{equation}
If the delta function is centred at a location $c$ then the effect of convolution is to translate the distribution $f(X)$ 
\begin{equation}
	f(X+c) = f(X) \ast \delta_c(X).
\end{equation}
There also exists the concept of a convolution power defined as 
\begin{equation}
	f^{\ast n} = \underbrace{f*f*\cdots*f*f}_n
\end{equation}	
and 
\begin{equation}
	f^{\ast 0} = \delta_0
\end{equation}
Given that appropriate conditions are met, the Central Limit Theorem says that a distribution function $f(X)$ with a mean of zero and a variance of $\sigma^2$ then $\lim x \to \infty\;\; x^{\ast n}/\sigma\sqrt{n}$ tends weakly towards the standard normal distribution.  Convolution of one distribution function with another can be considered as equivalent to diffusion.

\newcommand\ttop{4.0}

\begin{figure}[!ht]
	\begin{center}	
			\begin{tikzpicture}[scale=1] 
				\draw (-2,0) -- (+4,0); 
				\begin{axis}[%
					width=12cm,
					xlabel=$X$,
					ylabel=$F(X)$,
					grid=major,
					domain=-10:10,
					ymin=0,
					ymax=1,
					xmin=-10,
					xmax=10,
					anchor=origin,
					% tell pgfplots to place its anchor at (0,0):
					% (This is actually the default and can be omitted)
					at={(0pt,0pt)},
					% tell pgfplots to use the "natural" dimensions:
					disabledatascaling,
					% tell pgfplots
					% legend entries={gnuplot, Bowling et al},
					% legend pos=south east]
					]
					\addplot [smooth, black] {normcdf(x,0,3)};
					\addplot [dotted, semithick, black] {normcdf(x,0,9)};
					\addplot [smooth, thin, orange] {normcdf(x,0,9.48683)};
					
					\draw[black,fill=black] (-2,0.252493) circle (0.3ex);
					\draw[black,fill=black] (-6,0.252493) circle (0.3ex);
					\draw[black] (-2,0.252493) -- (-6,0.252493) node[midway,below] {$a(-2,t)\Delta t$};
					\coordinate (insetPositionZ) at (-6.9,0.252493);
					
					\draw[black,fill=black] (-1,0.369441) circle (0.3ex);
					\draw[black,fill=black] (-3,0.369441) circle (0.3ex);
					\draw[black] (-1,0.369441) -- (-3,0.369441) node[midway,below] {$a(-1,t)\Delta t$};
					\coordinate (insetPositionA) at (-3.9,0.369441);
					
					\draw[black,fill=black] (0,0.5) circle (0.3ex);
					\draw[black] (0,0.5) -- (0,0.5) node[midway,below] {$a(0,t)\Delta t$};
					\coordinate (insetPositionzero) at (-0.9,0.5);
									
					\draw[black,fill=black] (1,0.630559) circle (0.3ex);
					\draw[black,fill=black] (3,0.630559) circle (0.3ex);
					\draw[black] (1,0.630559) -- (3,0.630559) node[midway,below] {$a(1,t)\Delta t$};
					\coordinate (insetPositionB) at (2.2,0.630559);
					
					\draw[black,fill=black] (2,0.747) circle (0.3ex);
					\draw[black,fill=black] (6,0.747) circle (0.3ex);
					\draw[black] (2,0.747) -- (6,0.747) node[midway,below] {$a(2,t)\Delta t$};
					\coordinate (insetPositionC) at (5.2,0.747);
					%\draw[-{circle (0.1ex) }] (2,0.75)  -- (6,0.75) node[midway,above] {$\aplus a(X=2,t)$}; 
				\end{axis}
				\begin{axis}[at={(insetPositionZ)}, 
					width=2.5cm,
					axis line style={draw=none}, 
					tick style={draw=none},
					xticklabels={,,},
					yticklabels={,,}
					]
					%small plot
					\addplot [fill=gray!20, 
					domain=-0.5:0.5
					] {gaussfill(0,0.2,0)} \closedcycle;								
				\end{axis}
				\begin{axis}[at={(insetPositionA)}, 
					width=2.5cm,
					axis line style={draw=none}, 
					tick style={draw=none},
					xticklabels={,,},
					yticklabels={,,}
					]
					%small plot
					\addplot [fill=gray!20, 
		          domain=-0.5:0.5
		          ] {gaussfill(0,0.2,0)} \closedcycle;								
				\end{axis}
				\begin{axis}[at={(insetPositionB)}, 
				width=2.5cm,
				axis line style={draw=none}, 
				tick style={draw=none},
				xticklabels={,,},
				yticklabels={,,}
				]
				%small plot
				\addplot [fill=gray!20, 
				domain=-0.5:0.5
				] {gaussfill(0,0.2,0)} \closedcycle;								
			\end{axis}
			\begin{axis}[at={(insetPositionC)}, 
				width=2.5cm,
				axis line style={draw=none}, 
				tick style={draw=none},
				xticklabels={,,},
				yticklabels={,,}
				]
				%small plot
				\addplot [fill=gray!20, 
				domain=-0.5:0.5
				] {gaussfill(0,0.2,0)} \closedcycle;								
			\end{axis}
			\begin{axis}[at={(insetPositionzero)}, 
				width=2.5cm,
				axis line style={draw=none}, 
				tick style={draw=none},
				xticklabels={,,},
				yticklabels={,,}
				]
				%small plot
				\addplot [fill=gray!20, 
				domain=-0.5:0.5
				] {gaussfill(0,0.2,0)} \closedcycle;								
			\end{axis}
			\begin{scope}
				%\draw[dashed] (-1,2) -- (-1,-1);  65333333333333333333333333				%\node[pin=140:first] at (0,0) {};
				%\node[pin=second] at (3,2) {};
				%\node[pin=45:third] at (7,1) {};
				%\node[pin=0:fourth] at (5,-2) {};
			\end{scope}
				%\draw[dashed] (-1,2) -- (-1,0);  \node [below] at (0,0) {$t_0$};
				%\draw[dashed] (1,2) -- (1,0);  \node [below] at (1,0) {$t_1$};
				%\draw[dashed] (1,2) -- (1,0);  \node [below] at (1,0) {$t_1$};
				%\node [left] at (0,0.5) {X};  \draw[black,fill=black] (0,0.5) circle (0.1ex);
				%\draw[-{Latex[length=3mm, width=2mm]}]  (0,0.5) -- (\ttop,0.75) node[midway,above] {$p=1$}; 
				%\node [right] at (\ttop,0.75) {$X+\Delta X$}; \draw[black,fill=black] (\ttop,0.75) circle (0.1ex);
				%\node [below] at (\thalf,-0.2) {(a) Deterministic};
			\end{tikzpicture}
		\caption{Visualisation of a drift-first process increment.  Starting with a Normal CDF  ${\cal N}(0,3)$ (the solid black line) and having a drift increment of $a(X,t)=2X$ and $\Delta t = 1$, the drifted distribution becomes  ${\cal N}(0,9)$ (the dotted black line). Subsequently the diffusion operation is applied to mass at each point of the drifted distribution, represented by the small grey bell shaped functions, and masses are redistributed around the locus of the point of the mass by a convolution operation according to the form of the stochastic increment. Here the orange line is the dotted line convolved with a Normal Distribution  ${\cal N}(0,3)$, and represents the result of the drift-first process increment on the original distribution. \label{fig:visualisedrift}}
	\end{center}	
\end{figure}
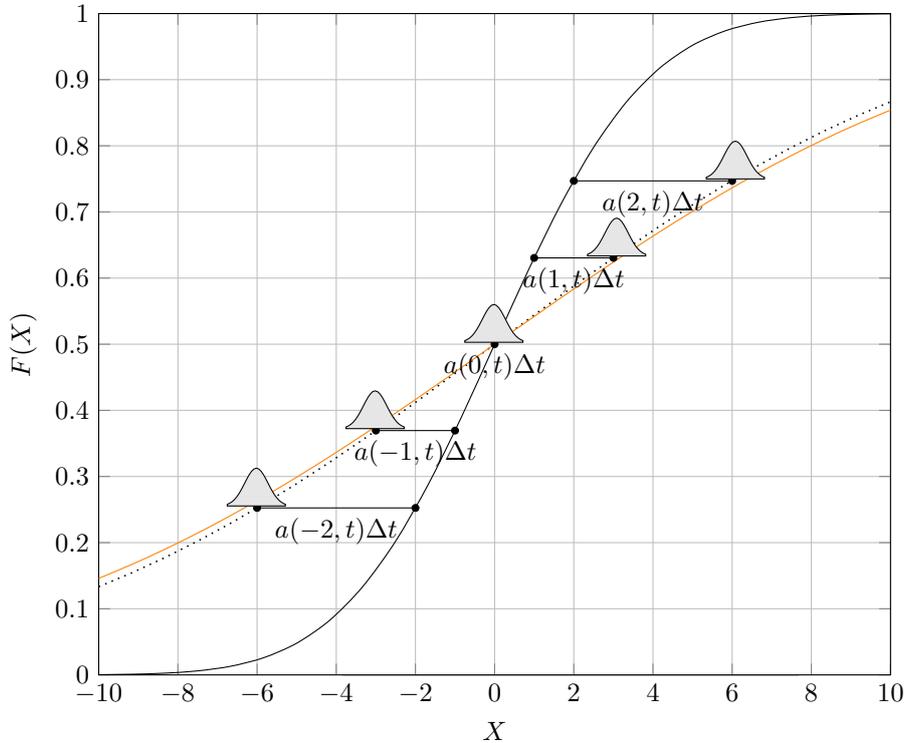

\subsection{A Visualisation of Deterministic and Diffusion Transformations}

The figure \ref{fig:visualisedrift} shows a cumulative distribution function undergoing a deterministic transformation, being stretched by an increment of $a(X,t)  \Delta t= 2X_i \Delta t$, followed by a diffusion transformation.  The combination of the drift (deterministic) and diffusion transformations is a single step in a standard Moate Simulation.  The example purposely introduces a drift term which is spatially dependent rather than a simple constant drift.  This drift is also large to enable the drifted loci to be easily seen and annotated.  The widths of the overlying  grey Gaussian diffusion functions shown are illustrative and not accurate.  It is clear from the diagram that the diffusion applied to the dotted line results in a wider CDF, as shown with the orange line.

\section{Moate Simulation\label{sec:MoateSimulation}}
We have just introduced the important transformations that allow us to cast a generalised It\^o process which is continuous in time and space
\begin{equation}
	dX = a(X,t) \;dt  + b(X,t)\; dW.
\end{equation}
into its discrete time/continuous space equivalent
\begin{equation}
	\Delta X = a(X,t) \;\Delta t  + b(X,t)\; \Delta W \label{eqn:discretetimeItoProcess}
\end{equation}
and this represents a single step in a multi-step Moate Simulation.  

This allows us to consider the generalised  It\^{o} process as a transformation that changes one distribution $f^{(t)}(X)\equiv f(X,t)$ into another $f^{({t+\Delta t})} \equiv f(X,t+\Delta t)$ through a drift with coefficient $a_t \equiv a(X,t)$ and a diffusion with coefficient $b_t \equiv b(X,t)$ that multiplies the normal distribution with variance $\Delta t$ given as $\Delta W ={\cal N}(0,\Delta t) \equiv {\cal N}$; the discrete time equivalent can be written as a drift-first approximation
\begin{equation}
	f^{(t +\Delta t)} =  \left(f^{(t)} \aplus  a_t \;\Delta t\right)  \ast  \left(b_t\; {\cal N}\right) \label{eqn:driftfirst}
\end{equation}
or a diffusion-first approximation.
\begin{equation}
	f^{(t +\Delta t)} =   \left(f^{(t)}  \ast  (b_t\; {\cal N}) \right) \aplus  a_t \;\Delta t \label{eqn:diffusionfirst}
\end{equation}	
The two approximations are not identical in the general case but where the drift $a_t$ is independent of location $X$ the two operations yield the same outcome.

Consider a multi step Moate Simulation, starting with a distribution $f^{(0)}$.   The parenthesised superscript on $f$ indicates the number of $\Delta t$ sized steps taken.  The drift-first step at $t_0 +\Delta t$ yields $f^{(1)}$ as
\begin{equation}
	f^{(1)} = \left(f^{(0)} \aplus  a_0 \Delta t\right)  \ast  \left(b_0\; {\cal N}\right)
\end{equation}
and after the second step $f^{(2)}$ is given by
\begin{eqnarray}
	f^{(2)} &=& \left(f^{(1)} \aplus  a_1 \Delta t\right)  \ast  \left(b_1\; {\cal N}\right) \\
	    &=& \left(\left( \left(f^{(0)} \aplus  a_0 \Delta t\right)  \ast  \left(b_0\; {\cal N}\right)\right) \aplus  a_1 \Delta t \right)  \ast  \left(b_1\; {\cal N}\right)
\end{eqnarray}	  
so that the process is described as a set of nested operations with the recursion relationship
\begin{equation}
	f^{(n)}  = \left(f^{(n-1)} \aplus  a_{n-1}\Delta t \right)  \ast  \left(b_{n-1}\; {\cal N}\right).
\end{equation}
Likewise, the diffusion-first recursion relationship is 
\begin{equation}
	f^{(n)}  =   \left(f^{(n-1)}  \ast  (b_{n-1}\; {\cal N}) \right) \aplus  a_{n-1}\Delta t.
\end{equation}	
%The evolution of the distribution function $f$ under a stochastic process is named {\bf Moate Simulation} to distinguish it from the existing Monte Carlo and  Binomial or Trinomial Tree approaches. 
It is possible in some special cases to obtain an analytic expression for $f^{(n)}$\label{point:constcoeff}. This is possible
where the process is purely deterministic, or where the drift $a_t$ and diffusion $b_t$ coefficients are constants.  In the latter case, the result is
\begin{equation}
	f^{(n)} = f^{(0)} \ast {\cal N}\left(n a\, \Delta t, n b^2\,\Delta t\right).
\end{equation}
However, for more general forms of $a$ and $b$ which may depend on both time $t$ and location $X$, analytic results may not easily present themselves. Fortunately, the numerical approach to Moate Simulation outlined in the next section can be tractable,  accurate and an efficient 
way to obtain the terminal or intermediate distributions, allowing for it to be utilised in many practical ways.

\section{A Numerical Approach to Moate Simulation}
Recalling the single step of a discrete time generalised It\^o process as found in equation \ref{eqn:discretetimeItoProcess}, if the diffusion coefficient is not a constant then the rate of diffusion from each locus of the probability distribution will be different and this presents an `apparent' impediment to the direct use of convolution in Moate Simulation.   The remedy for this is to simulate a different stochastic variable which is derived from the stochastic differential equation of the original stochastic variable but has a constant diffusion term for all loci, and at the end of the simulation transform back to the original stochastic variable.

\subsection{Variable Change to Constant Diffusion}
The first step is to change the variable using It\^o-Doeblin formula so that the diffusion coefficient in the transformed variable is a constant, if this is not already the case.  The drift coefficient will change too and may not be a constant but this is not in itself an impediment to obtaining a numerical result.

For example, the classic case is the Samuelson process 
\begin{equation}
	dS_t = \mu S_t \; dt \; + \; \sigma S_t \; dW_t  \label{eqn:Samuelson}
\end{equation}
where the price $S_t$ is transformed to $L_t$ where $L_t = \log S_t$.
\begin{eqnarray}
	dL_t &=&  \frac{\partial L_t}{\partial S_t} dS_t  +\frac{1}{2} \frac{\partial^2 L_t}{\partial S_t^2} d\langle S\rangle_t \nonumber\\
	     &=&\frac{1}{S_t}\left( \mu S_t \; dt \; + \; \sigma S_t \; dW_t \right) +\frac{1}{2} \left( \frac{-1}{S_t^2}\right) \sigma^2 S_t^2 \;dt \nonumber\\
	     &=& (\mu - \frac{1}{2}\sigma^2) \;dt \; + \; \sigma dW_t \label{eqn:SamuelsonL}
\end{eqnarray}
The change of stochastic variable from $S_t$ to $L_t$ has produced an expression in equation \ref{eqn:SamuelsonL} where the coefficient of the Wiener increment is the constant $\sigma$.  
The Moate Simulation can be used here with $L_0 = \log S_0$ is the point at time $t=0$ where the initial distribution is known.  If we assume that the initial PDF is represented by a Dirac $\delta$ function, or equivalently the initial CDF is represented by a unit step (Heaviside) function, then distribution after the first discrete step $\Delta t$ is given by
\begin{eqnarray}
f_{\Delta t}(L,\Delta t) &=& L_0 + (\mu - \frac{1}{2}\sigma^2) \Delta t  +  \sigma {\cal N}(0,\Delta t) \\
                &=& {\cal N}\left(L_0 + (\mu - \frac{1}{2}\sigma^2) \Delta t, \sigma^2 \Delta t\right)
\end{eqnarray}
This case is straightforward and, as discussed above, gives an analytic distribution at some final time $T$ of 
\begin{equation}
f_{T}(L,T)  = {\cal N}\left(L_0 + (\mu - \frac{1}{2}\sigma^2) T, \sigma^2 T\right).
\end{equation}
so a numerical approach is not necessary here.  

Consider a different case where even after transformation the diffusion coefficient is constant in time and space but drift term is not.  The process below contains a drift term in the square of $S_t$.
\begin{equation}
	dS_t = \mu S_t^2 \; dt \; + \; \sigma S_t \; dW_t
\end{equation}
Applying It\^o's lemma to this equation, with  $L_t = \log S_t$ as before, results in 
\begin{eqnarray}
	dL_t &=& (S_t\mu - \frac{1}{2}\sigma^2) \;dt \; + \; \sigma dW_t \nonumber\\
	     &=& (\exp(L_t) \mu - \frac{1}{2}\sigma^2) \;dt \; + \; \sigma dW_t \label{eqn:SamuelsonL2}
\end{eqnarray}
Another case might be to have no $S_t$ in the drift coefficient so that
\begin{equation}
	dS_t = \mu  \; dt \; + \; \sigma S_t \; dW_t
\end{equation}
so that
\begin{equation}
		dL_t = (\mu \exp(-L_t) - \frac{1}{2}\sigma^2) \;dt \; + \; \sigma dW_t 
\end{equation}
A third case involves the transformation of variable $M_t = S_t^2$ in the equation
\begin{equation}
	dS_t = \mu  S_t^{-1} \; dt \; + \; \sigma S_t^{-1} \; dW_t
\end{equation}
so that after application of It\^o's lemma
\begin{equation}
	dM_t = (2 \mu - \frac{\sigma^2}{M_t}) \;dt \; + \; 2 \sigma\; dW_t 
\end{equation}
None of these latter three stochastic differential equations are proposed for any particular physical or financial purpose; they merely demonstrate that a change of stochastic variable can lead to the coefficient of the Wiener increment being a constant.   Note in the third case the transformed drift coefficient will be undefined when $M_t = 0$ so the domain of the stochastic variable should be carefully considered.  In this case the original stochastic differential would be have been undefined at $S_t=0$ too so any modelling difficulties arising from this should have been recognised at this earlier stage.

In the case where we start with a discrete problem, we can approximate it as being continuous, transform the variable with It\^o's lemma to the required form, and then re-discretise the transformed stochastic differential equation.

Given the spatially constant form of the diffusion coefficient the evolution of the probability distribution can be attempted numerically, assuming that the continuous stochastic differential equation can be approximated in discrete time and that the distribution can be adequately represented on a grid.  

\subsection{Simulation}
Suppose that we have transformed to a random variable $X$ that gives us a drift-first approximation, with a constant drift term $c = b(X,t)$ then the distribution $f$ will evolve as 
\begin{equation}
		f(X,t+\Delta t) =   \left(f(X,t) \aplus  a(X,t) \;\Delta t\right)  \ast  \left(c\; {\cal N}(0,\Delta t)\right) \label{eqn:driftfirstdiscrete}
\end{equation}
The distribution in the initial state is known at a given time $t_0$, and is denoted as $f(X,t_0)$.   We can choose to work with $f(X,t)$, the cumulative distribution function (CDF), or $F(X,t_0)$, its equivalent probability density function (PDF), although there are pros and cons numerically with either functional form.

Where the location at $t_0$ is at a single discrete point the PDF can be represented as a Dirac delta function $\delta(X_0)$ or its cumulative equivalent, and Heaviside function $H(X_0)$.  The first step in a drift-first simulation will shift the delta function to $\delta(X_0 + a(X,t_0))$ and the diffusion step is applied to this so that the distribution at $t_0 +\Delta t$ is given by ${\cal N}(X_0 + a(X,t_0)\Delta t, c^2 \Delta t )$ and so the first function we need to represent on the grid is the distribution after the first step.  This distribution function can be represented on a discrete grid.

It is also possible to start with a known distribution at $t_0$ and represent it directly onto a discrete grid.  Since we want to apply drift to the distribution and then convolve it with $c {\cal N}(0,\Delta t)$  (or vice-versa) we need a grid representation which will represent the distribution at $t$, the diffusion function $c {\cal N}(0,\Delta t)$, and the next distribution at $t+\Delta t$ sufficiently accurately. The extent of the grid should be selected so that the value of the probability density function is negligible outside the grid.  Many convolution algorithms using Discrete Fourier Transformations (DFT) require a uniformly spaced grid although Non-Uniform Discrete Fourier Transform (NUDFT) approaches also exist.  The use of Fourier Transform techniques, particularly the Discrete (or Fast) Fourier Transforms (DFT/FFT), are both highly efficient and readily available in many mathematical libraries, and these algorithms can use specialised hardware like vector, tensor or GPU processors for to greatly speed up computation.

We consider a function at time $t$ on a grid as a set of $n$ tuples of position $X_i$ and the value $f_i \equiv f(X_i,t)$, that is $(X^{(t)}_i,f^{(t)}_i) \forall i \in n$.  Drifting the function does not change the value of $f^{(t)}_i$ in the tuple but it does change the location $X^{(t)}_i$, so that $X^{(t)}_i$ is replaced with $X^{(t)}_i +a(X^{(t)}_i,t) \Delta t$.   The drifting, in effect, creates a distortion of the location element of the tuple.   Since for the convolution operation a uniform grid is required, the set of $n$ tuples needs to be interpolated on to a suitable uniform grid (here $m$ tuples indexed by $j$).  The interpolation function selected should depend on the functional form of the distribution but it may be convenient to use  Piecewise Cubic Hermite Interpolating Polynomial (PCHIP)  interpolation or cubic spline interpolation.
%given the associations between Hermite Polynomials, normal distributions and Fourier Transforms.  

So the drift step is followed by interpolation onto a new uniform grid, and then the convolution operation is applied with the diffusion function $c {\cal N}(0,\Delta t)$.  The drift-first simulation step is summarised as the sequence of operations
\begin{equation}
    \{(X^{(t)}_i,f^{(t)}_i)\} \;\; 
    \underrightarrow{\hbox{drift\ }} \;\; \{(X^{(t)}_i + a(X^{(t)}_i,t) \Delta t, f^{(t)}_i)\} \;\; 
    \underrightarrow{\hbox{interpolate\ }} \;\;  \{(X^{(t+\Delta t)}_j,f^{(t)}_j)\}  \;\; 
    \underrightarrow{\hbox{convolve\ }}  \;\;  \{(X^{(t+\Delta t)}_j,f^{(t+\Delta t)}_j)\}.\nonumber
\end{equation}
The equivalent diffusion-first operation sequence is 
\begin{equation}
	\{(X^{(t)}_i,f^{(t)}_i)\} \;\; 
	\underrightarrow{\hbox{convolve\ }}  \;\;  \{(X^{(t)}_i,f^{(t+\Delta t)}_i)\}\;\; 
	\underrightarrow{\hbox{drift\ }} \;\; \{(X^{(t)}_i + a(X^{(t)}_i,t) \Delta t, f^{(t+\Delta t)}_i)\} \;\; 
	\underrightarrow{\hbox{interpolate\ }} \;\;  \{(X^{(t+\Delta t)}_j,f^{(t+\Delta t)}_j)\}.  \nonumber
\end{equation}
We should expect that as the time step $\Delta t$ is reduced towards zero the distribution produced by drift-first and diffusion-first respective simulations converge.  The degree to which these distributions differ can be used  to create measure of numerical error arising from the choice of drift or diffusion first sequence.  
%It may also be possible to use Richardson Extrapolation on the time increment to determine distribution $f(X,t)$ at the $\lim \Delta t \rightarrow 0$ corresponding to the distribution arising from the equivalent continuous time process.

There are a number of technical details associated with the convolution step that may merit a mention. Firstly, DFT algorithms for convolution often require padding and for a PDF this is normally with zeros, while for a CDF it should be left padded with zeros and right padded with ones.  This means that for the PDF a renormalisation step may be useful if the numerical integration of the PDF does not yield hard unity;  without this renormalisation multiple simulation steps may introduce a systematic bias which affects the accuracy of integrals involving the PDF.   The use of the CDF during the simulation ensures that the integration of its corresponding PDF will be unity by construction, but particular care needs to be taken to prevent distortion at the right edge of the distribution as some algorithms will attempt to pad on the right with zeros that can cause the right tail of the CDF to dip away from unity.  

Additionally, Fourier Transform based convolution algorithms may benefit from the recognition that the diffusion function, here $c {\cal N}(0,\Delta t)$, is identical on each step of the simulation so some numerical speed-up may be achieved by only having to transform the distribution function $f$ on each step and convolve this with a retained transformed diffusion function (assuming the grid spacing is not altered between simulation steps).

As diffusion occurs on each step of the simulation it may also be necessary to increase the extent and/or the spacing of the grid. Extending the grid occurs naturally by selecting a constant threshold for the tails of the CDF up to which the grid extends from the centre in both directions.  

%Thirdly, one of the features of a Fourier transformation on a Gaussian function is that the transformed function is also Gaussian. (INTERESTING??)

\subsection{Example: Vanilla Option Pricing using Moate Simulation}
We could, in principle, select any general It\^o process to demonstrate Moate Simulation but the choice of the Black-Scholes model allows for the comparison of the accuracy of the numerical results with analytic results.   
The Black-Scholes European call and put option pricing formulae for equities are based on the Samuelson process (equation \ref{eqn:Samuelson}) and these analytic formulae can be used to compare the option prices with those determined using Moate Simulation using the log process in equation \ref{eqn:SamuelsonL}.

The Black-Scholes European call option (no dividend) has a closed form analytic solution for its value $C_t$ given by
\begin{equation}
	C_t  = S_t \Phi \left(\frac{\log\frac{S_t}{K} + (r +\frac{1}{2} \sigma^2) \tau}{\sigma\sqrt{\tau}}\right) - K e^{-r\tau}  \Phi\left(\frac{\log\frac{S_t}{K} + (r -\frac{1}{2} \sigma^2) \tau}{\sigma\sqrt{\tau}}\right)  
\end{equation}
while the corresponding European put option value, $P_t$, is given by
\begin{equation}
	P_t  =
	K e^{-r\tau}  \Phi \left(\frac{\log\frac{S_t}{K} + (r -\frac{1}{2} \sigma^2) \tau}{\sigma\sqrt{\tau}}\right)  
	 - S_t \Phi \left(-\frac{\log\frac{S_t}{K} + (r +\frac{1}{2} \sigma^2) \tau}{\sigma\sqrt{\tau}}\right) 
\end{equation}
where   $S_t$ is the current spot price, $K$ is the strike price, $r$ is the risk-free rate, $\sigma$ is the volatility, $\tau$ is the time until the option expiry, and $\Phi(\cdot)$ is the standard normal cumulative distribution function.

\begin{table}[h]	
	\begin{center}
		\begin{tabular}{|c|l|c|l|c|r|}
			\hline\hline
			Log Spacing & Call Price & Abs diff & Put Price & Abs diff & Run-time (sec)   \\\hline\hline
			Analytic & 0.120165592579702 & --- & 0.210452117932772 & --- & $\sim$2e-3 \\\hline
			1E-05 & 0.120165591 & 9E-10 & 0.2104521175 & 4E-10 & 26.92                \\
			5E-05 & 0.120165595 & 2E-09 & 0.210452119 & 2E-09 & 3.95                  \\
			1E-04 & 0.1201656 & 1E-08 & 0.21045212 & 8E-09 & 2.28                     \\
			5E-04 & 0.1201659 & 3E-07 & 0.2104523 & 2E-07 & 1.03                      \\
			1E-03 & 0.120166 & 1E-06 & 0.2104529 & 9E-07 & 0.94                       \\
			5E-03 & 0.1202 & 3E-05 & 0.21047 & 2E-05 & 0.80                            \\\hline
			Monte Carlo 25mm & 0.12015 & 9E-06 & 0.21044 & 1E-05 & 182.98       \\ \hline\hline
		\end{tabular}
		\caption{Moate Simulation results for Black-Scholes call and put option for daily (365 iterations) grid evolution. The vanilla European options have a strike of $K=4.30$, expiration $\tau=1$, risk-free rate $r=0.05$, volatility $\sigma=0.1$ and the current spot price $S_t=4.00$. 
			The precision shown is to the first differing digit from the analytic value, and the relative difference is the absolute difference to the analytic value. Probability distributions are retained to a CDF threshold of  1e-12. A Monte Carlo calculation using 25 million paths and antithetic sampling provides a comparator to the Moate Simulation results. The run-times shown for each calculation using an AMD Phenom II X6 1075T 3.00 GHz Processor (2010) are for very rough comparison only and negligible optimisation of the code was attempted for the simulations.\label{tab:bsresults}}
	\end{center}
\end{table}

\begin{table}[h]	
	\begin{center}
		\begin{tabular}{|c|c|c|c|c|c|c|c|c|}	
			\hline\hline        
			Log Grid& Start  & Start    & Start     & End       & End      & End      & End   & End    \\
			spacing & min L  & max L    & grid size & grid size & min L    & max L    & Min S & Max S   \\\hline\hline
			1E-05 & 1.353138 & 1.419698 &      6657 &    122453 & 0.819034 & 2.043554 &  2.27 & 7.72   \\
			5E-05 & 1.353168 & 1.419668 &      1331 &     24491 & 0.819044 & 2.043544 &  2.27 & 7.72  \\
			1E-04 & 1.353218 & 1.419618 &       665 &     12245 & 0.819094 & 2.043494 &  2.27 & 7.72  \\
			5E-04 & 1.353418 & 1.419418 &       133 &      2449 & 0.819294 & 2.043294 &  2.27 & 7.72  \\
			1E-03 & 1.353418 & 1.419418 &        67 &      1225 & 0.819294 & 2.043294 &  2.27 & 7.72  \\
			5E-03 & 1.356418 & 1.416418 &        13 &       245 & 0.821294 & 2.041294 &  2.27 & 7.70  \\ \hline\hline	
		\end{tabular}                                               
		\caption{Moate Simulation numerical details for the same options presented in table \ref{tab:bsresults}. For the calculation 365 daily iteration steps were used for differing uniform grids in the Log Price space.  Probability distributions were retained to a CDF threshold of  1e-12 at both ends of the distribution.\label{tab:technicaldetails} }
	\end{center}
\end{table}

\begin{table}[h]
	\begin{center}
		\begin{tabular}{|c|c|l|c|l|c|}
			\hline\hline
			Source & Spacing & Call  & Call & Put  &  Put                                     \\
			&  & Price & Abs diff & Price & Abs diff                               \\\hline\hline
			analytic & analytic & 0.120165592579702 & --- & 0.210452117932772 & --- \\\hline
			365 & 1E-05 & 0.120165591 & 9E-10 & 0.2104521175 & 4E-10                \\
			1 & 1E-05 & 0.1201655923 & 2E-10 & 0.2104521178 & 7E-11                 \\
			pdf on grid & 1E-05 & 0.1201655924 & 8E-11 & 0.2104521178 & 8E-11       \\\hline
			365 & 5E-05 & 0.120165595 & 2E-09 & 0.210452119 & 2E-09                 \\
			1 & 5E-05 & 0.120165595 & 3E-09 & 0.210452119 & 2E-09                   \\
			pdf on grid & 5E-05 & 0.120165594 & 1E-09 & 0.210452119 & 1E-09         \\\hline
			365 & 1E-04 & 0.1201656 & 1E-08 & 0.21045212 & 8E-09                    \\
			1 & 1E-04 & 0.1201656 & 1E-08 & 0.21045212 & 9E-09                      \\
			pdf on grid & 1E-04 & 0.1201655923 & 2E-10 & 0.2104521177 & 2E-10       \\\hline\hline
		\end{tabular}
		\caption{Black-Scholes results with the same parameters as used in table \ref{tab:bsresults} that  compare analytic price with Moate Simulation prices using 365 daily steps or a single annual step (1), and with the numerically integrate analytic terminal probability density (pdf on grid) for the finer grid spacings.  Similar accuracy is obtained for a given spacing across all three integrations calculations.\label{tab:numericcomparison}}
	\end{center}
\end{table}

Consider the European call and put options where the Strike $K=4.30$, expiration $\tau=1$, risk-free rate $r=0.05$, volatility $\sigma=0.1$ and the current price $S_t=4.00$; using these parameters the call price $C_t=0.12016559257970$ and the put price $P_t= 0.21045211793277$ as calculated using the analytic expressions above.

% 0.1201565776558197  0.0003709416201964277      0.21044216005558877  0.00029161590234099343

An alternative approach is to use a simple Monte Carlo Simulation with antithetic sampling using 365 steps and 25 million paths for the same parameters; this provided results for the call and put respectively of  0.12015 %9.E-06 
and 0.21044 and these differ from the analytic values by about 1E-5. %with a standard deviation of approximately $3\times10^{-4}$.
While there are other refinements of Monte Carlo like the use of quasi-random sequences which could be employed to potentially improve the accuracy of the result, 25 million paths might be considered to be a large number of paths in practice.

We compare the analytic and Monte Carlo results to the Moate Simulation results.  Using different spacings on the log price grid the distributions of prices at the expiry can be obtained, and  the discounted expectation of the payoff profile $Y(S)$ assuming the time $T$ terminal distribution $f_T(S)$ gives the price according to the equation
\begin{equation}
	V_t =  \exp(-r \tau) E_t^{\amsbb Q} [ Y(S)] 
\end{equation}
which is equivalently written as
\begin{equation}
	V_t = \exp(-r \tau) \int Y(S) f_T(S) \; dS.
\end{equation}
For call options $Y(S)= \max(0,S_t-K) = (S_t-K)^+$ while for put options $Y(S)= \max(0,K- S_t) = (K -S_t)^+$.

The cumulative probability distribution function is calculated as $f_i = \sum_{j=0}^i F_j$ or otherwise estimated from the probability density function $F$. 
The distribution at expiry is given for $L_T = \log S_T$ on the last step of the simulation. Transforming the variable from log price back to price is simple 
${(\exp(L_{i,T}), f_{i,T})} =   {((S_{i,T}), f_{i,T})} $   
and the PDF values on the new are obtained by differentiation of the CDF $f_T(X)$.
A number of different integration 
approaches\footnote{ The integration approaches tested included (i) simple integration on the log price grid,
(ii) simple integration on the regular price grid,
(iii) cubic spline interpolated function to perform  Romberg quadrature,
(iv) cubic spline interpolated function to perform Gaussian quadrature,
(v) PCHIP interpolated function to perform Romberg quadrature,
(vi) PCHIP interpolated function to perform Gaussian quadrature,
(vii) Simpson's rule.}
were compared but the accuracy of the integration was very similar across all approaches so the results presented here are the simple integration on the log price grid.

Table \ref{tab:bsresults} compares the analytic values with their corresponding Monte Carlo results, along with the Moate Simulation derived prices using different spacings on the Log price grid. The Monte Carlo results are comparable in accuracy to the coarsest grid (5E-03) results for Moate Simulation, although the Moate Simulation was around 1000 times faster.  As might be anticipated, finer the log price grid the more accurate the prices become; this continues are until the relative differences in the price appear to bottom out at around 8 or 9 significant figures of agreement.  This represents a large improvement on the Monte Carlo derived value in terms of both accuracy and speed.

Table \ref{tab:technicaldetails} shows the sizes of the grids that were used at the first step and last step of the simulation.   The distributions are bounded by a threshold which is set to be suitably low, in this case where the CDF is bounded at values of 1E-12 and 1.0 - 1E-12. As a result, the grid sizes are roughly proportional to the reciprocal of the grid spacing. As the distribution becomes more diffuse the grid size increases and, at the end of the simulation,  the number of grid points is close to 20 times larger than at the beginning.

To determine the extent to which the accuracy is affected by the initialisation or number of iterative steps, the numerical results are compared in table \ref{tab:numericcomparison}  where the analytic option prices are compared with a 365 daily step simulation, with a single (1) step simulation, and with the known analytic terminal distribution discretised (PDF) on the grid and integrated with the option payoff profiles in the same manner as the 365 day and 1 year results.  The results suggest that the a similar accuracy is achieved with all the numerical approaches (365,1,pdf on grid) and that the differences between these results and the analytic values is small with 8 to 9 significant figure agreement produced.  The remaining error is common to all three numerical approaches, and could arise from various causes including the truncation of the grids or the specific integration techniques used.

Analytic results will always be preferred where they exist, and Monte Carlo approaches may have unique advantages (e.g.\ for options with look back or averaging features),  but Moate Simulation holds out the exciting prospect of accurate pricing where the chosen stochastic processes do not have any known analytic solution.  This opens a significant new front in the modelling of financial, economic and other scientific domains which are built on top of stochastic processes.

% $\text{FinancialDerivative}[\{\text{European},\text{Call}\},\{\text{StrikePrice}\to 4.3,\text{Expiration}\to 1\},\{\text{InterestRate}\to 0.05,\text{Volatility}\to 0.1,\text{CurrentPrice}\to 4.,\text{Dividend}\to 0.\}]= 0.1201655925797023$

%FinancialDerivative[{"European", "Call"}, {"StrikePrice" -> 4.30, 
	%	"Expiration" -> 1},  {"InterestRate" -> 0.05, "Volatility" -> 0.1, 
	%	"CurrentPrice" -> 4.0, "Dividend" -> 0.0}] 
% 0.1201655925797023

%$\text{FinancialDerivative}[\{\text{European},\text{Put}\},\{\text{StrikePrice}\to 4.3,\text{Expiration}\to 1\},\{\text{InterestRate}\to 0.05,\text{Volatility}\to 0.1,\text{CurrentPrice}\to 4.,\text{Dividend}\to 0.\}] = 0.2104521179327725$
%FinancialDerivative[{"European", "Put"}, {"StrikePrice" -> 4.30, 
	%	"Expiration" -> 1},  {"InterestRate" -> 0.05, "Volatility" -> 0.1, 
	%	"CurrentPrice" -> 4.0, "Dividend" -> 0.0}]
% 0.2104521179327725

\section{Extensions}
We demonstrated in the last section that the Moate Simulation approach can produce results of high numerical accuracy.  In this section we consider a set of extensions which address derivative pricing features that are of interest to some practitioners in finance. 

\subsection{Altered Payoff Profiles}
The payoff profile $Y(S_T)$ was stated above for vanilla European call and put options as $\max(0,S_T-K)$ and $\max(0,K-S_T)$ respectively.  If an arbitrary payoff profile is sufficiently accurately represented on the selected grid then it is possible to price the derivative.  An example of a digital (or binary) payoff profile pays 1 unit at $T$ if  $S_T >K$ otherwise pays nothing; this  payoff is a Heaviside step function $H(\cdot)$ so that  $Y(S_T)= H(K)  =  \mathbf{1}_{\{S_T>K\}}$.    For European power options the call and put payoff functions are $\max(0,S_T^\alpha-K)$ and $\max(0,K-S_T^\alpha)$ where the exponent $\alpha$ is greater than zero. The expectation here is that as long as both the payoff function $Y(S)$ and the distribution at expiry $f(S)$ are well represented on the grid then numerical integration should produce acceptably accurate derivative valuations.

\subsection{Dividends}
Dividends can be treated in a number of different ways.   Firstly, the payment may be modelled to pay out on a discrete dates, or be considered to pay continuously.  The former case is the typically observed market behaviour while the latter case is easier to model producing tractable analytic solutions starting with processes like the modification of the Samuelson Process in equation \ref{eqn:Samuelson} to include a continuous dividend rate (also called dividend yield) $D_y$ as 
\begin{equation}
	dS_t = (\mu -D_y)  S_t \; dt \; + \; \sigma S_t \; dW_t  \label{eqn:SamuelsonDividend}.
\end{equation}
Secondly, the dividend payment (or yield) is most commonly modelled as a known or predicted discrete value, but an alternative is to model the dividend payment or yield  as a distribution.

The four different cases from the permutations from discrete or continuous payment timing with the choice of discrete dividend value versus a distribution of values are all tractable with Moate Simulation.  The discrete payment on known dates (the typical market convention) is easily achieved by subtracting the dividend payout $d$ so that $\{(S^{(t)}_i,f^{(t)}_i)\} \;\; \underrightarrow{\hbox{pay dividend\ }}  \;\; 	\{(S^{(t)}_i -d_t),f^{(t)}_i)\}$ on the particular dividend date $t$.  The case of a known dividend yield paid continuously is easily performed by modifying the drift coefficient $a$ by subtraction of the dividend yield $D_y$ as inferred from the example process in equation \ref{eqn:SamuelsonDividend}.   Where the dividend payment is modelled as a distribution, this can be handled as an additional convolution step on the payment date.

\subsection{Barriers}
Barrier options, which knock-in or knock-out if the price is above (or alternatively below) some barrier $B_t$ on a given date $t$ are treated in Moate Simulation by restricting the price distribution that is carried over onto the next step of the simulation.  For example if the up-and-out style barrier option is designed to allow only the distribution $f(S_t)$ where $S_t<B_t$ to propagate to the next step, then zeroing $f_i^{(t)}$ for $S^{(t)}_i >= B_t$ will yield a distribution in which some of the distribution has been `cast away' so that the remaining distribution integrates to the probability of the option being live after the barrier date.   If the barrier exists for multiple dates, or even all of the dates, then the restriction on the distribution is applied on each relevant date.  The effect is for the distribution that goes over the barrier to leak away, with the remnant distribution being used for the pricing at option expiry.    Multiple barriers are also straightforward.

Where the drift and diffusion coefficients are constants (as mentioned in section \ref{point:constcoeff}), or where analytic forms for the distribution are otherwise available, the distribution at the barrier can be obtained in a single step, distribution outside the barrier cast away, and the remaining distribution can be evolved to the next barrier (or expiry) with a single step.

\subsection{Early Exercise}
Options where early exercise is allowed are tractable with Moate Simulation.   These include American and Bermudan options; for these options the modelling the decision to exercise early can depend upon the specific features of the option, like whether it pays dividends or not.  The intrinsic value $I$ of an American put option is given as 
\begin{equation}
	I = \max \{ 0, K -S_t\} \; \forall t \in (0,T)  = \max(0, K-S_0, K-S_1,\ldots, K-S_{T-1}, K-S_T)
\end{equation}
If the intrinsic value of the option is greater than the terminal value then the investor should consider the sale of the option.   Since the distribution functions $f^{(t)}$ are known for each date $t$ it is possible to calculate the intrinsic value.
A similar approach may be taken for Bermudan options where exercise is allowed on a smaller set of dates.

\begin{table}[t]
	\begin{center}
		\resizebox{\columnwidth}{!}{
			\begin{tabular}{|c|c|c|l|c|l|c|l|c|l|c|l|c|l|c|}
				\hline\hline
				Finite & Type & Log & Price & Abs & Delta & Abs & Gamma & Abs & Rho & Abs & Theta & Abs & Vega & Abs \\
				Step &   & Spacing &   & Diff &   & Diff &   & Diff &   & Diff &   & Diff &   & Diff \\ \hline\hline
				analytic & call &  --- & 0.12016559258 &  --- & 0.431244511793 &  --- & 0.982506747863 &  --- & 1.604812454593 &  --- & -0.158841162559 &  --- & 1.572010796581 &  --- \\
				\hline\hline
				1E-04 & call & 1E-05 & 0.120165591 & 1E-09 & 0.4312445118 & 5E-11 & 1.01 & 3E-02 & 1.60481247 & 2E-08 & -0.1588411 & 6E-08 & 1.5720107 & 5E-08 \\
				1E-03 & call & 1E-05 &  &  & 0.43124454 & 3E-08 & 0.98251 & 4E-06 & 1.604814 & 2E-06 &  &  & 1.57201 & 4E-06 \\
				1E-02 & call & 1E-05 &  &  & 0.431247 & 3E-06 & 0.9825 & 5E-05 & 1.605 & 2E-04 &  &  & 1.572 & 4E-04  \\ \hline
				1E-04 & call & 5E-05 & 0.120165593 & 5E-10 & 0.43124451 & 1E-09 & 0 & 6E-01 & 1.60481247 & 2E-08 & -0.1588409 & 3E-07 & 1.57202 & 7E-06  \\
				1E-03 & call & 5E-05 &  &  & 0.43124454 & 3E-08 & 0.98251 & 4E-06 & 1.604814 & 2E-06 &  &  & 1.57201 & 4E-06  \\
				1E-02 & call & 5E-05 &  &  & 0.431247 & 3E-06 & 0.9825 & 5E-05 & 1.605 & 2E-04 &  &  & 1.572 & 4E-04  \\ \hline
				1E-04 & call & 1E-04 & 0.120165591 & 1E-09 & 0.4312 & 8E-05 & 0.6 & 3E-01 & 1.60481247 & 2E-08 & -0.15884 & 3E-06 & 1.5721 & 5E-05  \\
				1E-03 & call & 1E-04 &  &  & 0.4312445 & 3E-08 & 0.99 & 7E-03 & 1.604814 & 2E-06 &  &  & 1.57201 & 4E-06  \\
				1E-02 & call & 1E-04 &  &  & 0.431247 & 3E-06 & 0.98248 & 2E-05 & 1.605 & 2E-04 &  &  & 1.572 & 4E-04  \\ \hline
				1E-04 & call & 5E-04 & 0.1201657 & 1E-07 & 0.4315 & 3E-04 & 0 & 1E+00 & 1.605 & 1E-03 & -0.1589 & 5E-05 & 1.5718 & 1E-04  \\
				1E-03 & call & 5E-04 &  &  & 0.43124448 & 3E-08 & 1 & 4E-01 & 1.604814 & 2E-06 &  &  & 1.5718 & 1E-04  \\
				1E-02 & call & 5E-04 &  &  & 0.431247 & 3E-06 & 0.9823 & 1E-04 & 1.605 & 2E-04 &  &  & 1.572 & 4E-04  \\ \hline
				1E-04 & call & 1E-03 & 0.120166 & 4E-07 & 0.431 & 7E-04 & 0 & 1E+00 & 1.601 & 3E-03 & -0.1587 & 1E-04 & 1.5722 & 3E-04  \\
				1E-03 & call & 1E-03 &  &  & 0.431 & 7E-04 & 0 & 1E+00 & 1.604815 & 3E-06 &  &  & 1.5722 & 3E-04  \\
				1E-02 & call & 1E-03 &  &  & 0.431247 & 3E-06 & 0.97 & 1E-02 & 1.605 & 2E-04 &  &  & 1.572 & 4E-04  \\ \hline
				1E-04 & call & 5E-03 & 0.12018 & 2E-05 & 0.431 & 7E-04 & 0 & 1E+00 & 1.601 & 3E-03 & -0.1587 & 1E-04 & 1.5721 & 1E-04  \\
				1E-03 & call & 5E-03 &  &  & 0.431 & 7E-04 & 0 & 1E+00 & 1.601 & 3E-03 &  &  & 1.5721 & 1E-04  \\
				1E-02 & call & 5E-03 &  &  & 0.43122 & 1E-05 & 0 & 8E-01 & 1.605 & 2E-04 &  &  & 1.5717 & 3E-04  \\ \hline\hline
				analytic & put &  --- & 0.210452117933 &  --- & -0.568755488207 &  --- & 0.982506747863 &  --- & -2.48547407076 &  --- & 0.045673163709 &  --- & 1.572010796581 &  --- \\ 
				\hline\hline
				1E-04 & put & 1E-05 & 0.210452117 & 5E-10 & -0.568755488 & 2E-10 & 1.01 & 3E-02 & -2.48547405 & 1E-08 & 0.0456732 & 6E-08 & 1.5720108 & 4E-08  \\
				1E-03 & put & 1E-05 &  &  & -0.56875545 & 3E-08 & 0.98251 & 4E-06 & -2.485472 & 2E-06 &  &  & 1.57201 & 4E-06  \\
				1E-02 & put & 1E-05 &  &  & -0.568752 & 3E-06 & 0.9825 & 5E-05 & -2.4853 & 1E-04 &  &  & 1.572 & 4E-04  \\ \hline
				1E-04 & put & 5E-05 & 0.210452118 & 1E-09 & -0.568755489 & 1E-09 & 0 & 6E-01 & -2.48547405 & 2E-08 & 0.0456734 & 3E-07 & 1.57202 & 7E-06  \\
				1E-03 & put & 5E-05 &  &  & -0.56875545 & 3E-08 & 0.98251 & 4E-06 & -2.485472 & 2E-06 &  &  & 1.57201 & 4E-06  \\
				1E-02 & put & 5E-05 &  &  & -0.568752 & 3E-06 & 0.9825 & 5E-05 & -2.4853 & 1E-04 &  &  & 1.572 & 4E-04  \\ \hline
				1E-04 & put & 1E-04 & 0.210452117 & 7E-10 & -0.5688 & 8E-05 & 0.6 & 3E-01 & -2.48547405 & 1E-08 & 0.04568 & 3E-06 & 1.5721 & 5E-05  \\
				1E-03 & put & 1E-04 &             &       & -0.5687555 & 3E-08 & 0.99 & 7E-03 & -2.485472 & 2E-06 &  &  & 1.57201 & 4E-06  \\
				1E-02 & put & 1E-04 &  &  & -0.568752 & 3E-06 & 0.98248 & 2E-05 & -2.4853 & 1E-04 &  &  & 1.572 & 4E-04  \\ \hline
				1E-04 & put & 5E-04 & 0.2104522 & 1E-07 & -0.5684 & 3E-04 & 0 & 1E+00 & -2.484 & 1E-03 & 0.0456 & 5E-05 & 1.5718 & 1E-04  \\
				1E-03 & put & 5E-04 &  &  & -0.56875551 & 3E-08 & 1 & 4E-01 & -2.485472 & 2E-06 &  &  & 1.5718 & 1E-04  \\
				1E-02 & put & 5E-04 &  &  & -0.568752 & 3E-06 & 0.9823 & 1E-04 & -2.4853 & 2E-04 &  &  & 1.572 & 4E-04  \\ \hline
				1E-04 & put & 1E-03 & 0.210453 & 4E-07 & -0.569 & 7E-04 & 0 & 1E+00 & -2.488 & 3E-03 & 0.0458 & 1E-04 & 1.5722 & 3E-04 \\
				1E-03 & put & 1E-03 &  &  & -0.569 & 7E-04 & 0 & 1E+00 & -2.485471 & 2E-06 &  &  & 1.5722 & 3E-04 \\
				1E-02 & put & 1E-03 &  &  & -0.568752 & 3E-06 & 0.97 & 1E-02 & -2.4853 & 2E-04 &  &  & 1.572 & 4E-04  \\ \hline
				1E-04 & put & 5E-03 & 0.21046 & 2E-05 & -0.569 & 7E-04 & 0 & 1E+00 & -2.488 & 3E-03 & 0.0458 & 1E-04 & 1.5721 & 1E-04  \\
				1E-03 & put & 5E-03 &  &  & -0.569 & 7E-04 & 0 & 1E+00 & -2.488 & 3E-03 &  &  & 1.5721 & 1E-04  \\
				1E-02 & put & 5E-03 &  &  & -0.56877 & 1E-05 & 0 & 8E-01 & -2.4852 & 2E-04 &  &  & 1.5717 & 3E-04 \\  
				\hline\hline
			\end{tabular}    
		}                                           
		\caption{Moate Simulation results for Black-Scholes call and put option Greeks for daily (365 iterations) grid evolution. The vanilla European options have a strike of $K=4.30$, expiration $\tau=1$, risk-free rate $r=0.05$, volatility $\sigma=0.1$ and the current spot price $S_t=4.00$. 
			The precision shown is to the first differing digit from the analytic value, and the absolute difference is the absolute difference to the analytic value. Probability distributions are retained to a CDF threshold of  1e-12. \label{tab:bsgreekresults}  }
	\end{center}
\end{table}

\subsection{Computing Greeks}

Numerically computing of sensitivities of the option prices to changes in parameters, known informally in finance as the Greeks, is possible with finite differencing techniques.  However, special attention is required to avoid unacceptable loss in numerical precision.  We have seen in the  examples presented in table \ref{tab:bsresults} that the prices can be obtained to relatively high degrees of accuracy.  Table \ref{tab:bsgreekresults} shows the degree of accuracy that was achieved on different distribution function grid spacings and for different finite differencing step sizes to yield the derivatives of the price with respect to the different parameters. Where $V_t$ represents the option price,  the table includes Delta ($\frac{dV_t}{dS_t}$), Gamma  ($\frac{d^2V_t}{dS_t^2}$), Rho ($\frac{dV_t}{dr}$), Theta ($-\frac{dV_t}{d\tau}$) and Vega ($\frac{dV_t}{d\sigma}$) sensitivities.  The trade-off between precision in the price and size of the finite step is evident, particularly for Gamma where differencing occurs twice to obtain the second derivative.

\subsection{Jump Increments}
Although this paper has focused on processes with Wiener increments $dW$, represented over a period $\Delta t$ by $\Delta W$, a standard normal distribution  ${\cal N}(0,\Delta t)$, it is also possible to model jumps increments.
Merton\cite{Merton1976JumpDiffusion,Merton1990} introduced jumps into price processes in combination with diffusion to form a jump-diffusion process.  There are a number of variations of formulation for the jump increment that satisfy different conditions.  We have not examined these here but note that the Poisson distributions, key to introducing the jump into the dynamics,  can be handled in Moate Simulation using the same convolution techniques as discussed above for the Wiener increments.

\subsection{Higher Dimensional Models}
In section \ref{sec:MoateSimulation} the Moate Simulation was presented for a one dimensional problem, but it is extensible to higher dimensions where the convolution of the probability density function with a diffusion density function is valid if the two diffusion functions are independent.  Note that this is not the same as requiring that the marginal densities of either the initial probability density function or the diffusion function are independent; this is not a requirement here.

The convolution of two independent $k$ dimensional joint probability density functions, $F$ and $G$, over a domain $D \subseteq {\mathbb R}^k$ is given by a multidimensional integral
\begin{equation}
	(F*G)({\mathbf x})  =\int_{D} F({\mathbf x}-{\mathbf y}) G({\mathbf y}) \; d^k {\mathbf y}
\end{equation}
where $\mathbf{x}$ represents a tuple $(x_1,x_2,\cdots,x_k)$ and  ${\mathbf x}-{\mathbf y}$ represents $(x_1 -y_1,x_2-y_2,\cdots,x_k-y_k$).
If each of the marginal densities, like $F(x_1)$, is independent in joint distribution $F({\mathbf x})$ then 
\begin{equation}
	F({\mathbf x}) = \prod_j^k F_j(x_j)
\end{equation}
otherwise the marginal density needs to be determined by integrating the joint distribution over all the other variables, for example by 
\begin{equation}
	F(x_1) = \int\cdots\int F({\mathbf x}) \; dx_2 \cdots dx_k
\end{equation}

\begin{figure}[t]
	\begin{center}
		\begin{tikzpicture}
			\newcommand\Square[1]{+(-#1,-#1) rectangle +(#1,#1)}
			\draw[style={->},line width=0.5mm] (0,-4) -- (0,0);  \node [right] at (0,-2) {$a_2 \Delta t$};
			\draw[style={->},line width=0.5mm] (-6,-4) -- (0,-4);  \node [below] at (-3,-4) {$a_1 \Delta t$};
			\node [text width=5cm,minimum height=6cm,minimum width=10cm]  at (-3,1) {Diffusion density multivariate normal distribution with\\ correlation $\rho=0.75$};
			\draw[line width=0.2mm] (-6,-4) \Square{3pt};
			\node[opacity=0.6,inner sep=0pt] (image) at (0,0) {
				\includegraphics[trim={2cm 2cm 2cm 2cm},clip=true]{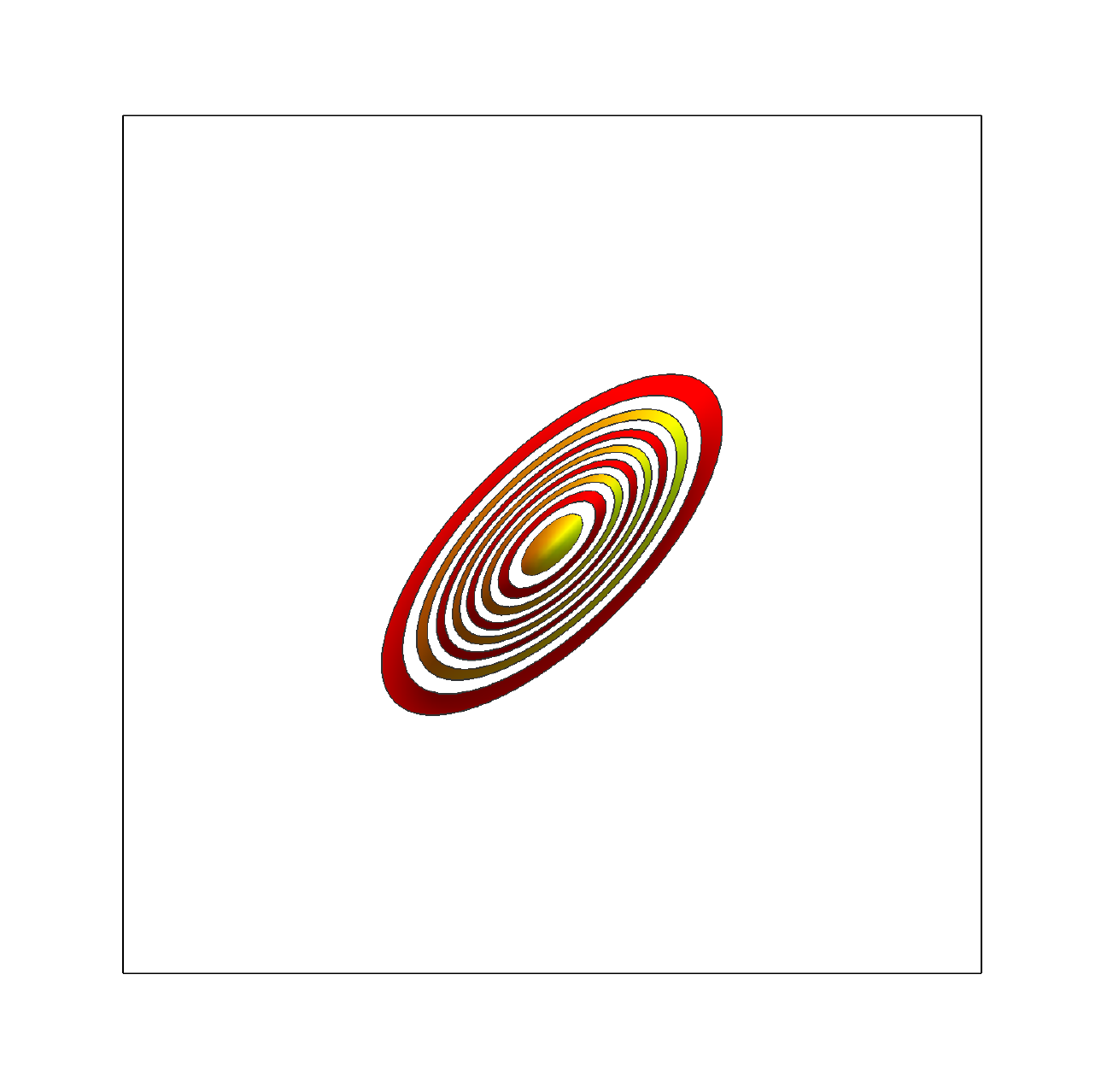}
			};
			
		\end{tikzpicture}
		\caption{Two dimensional Moate Simulation single step showing drift followed by diffusion with a multivariate normal density function with correlation of $\rho=0.75$.  The mass of the starting distribution around a single point, represented by the tiny square, is drifted and then diffused. The diffusion is independent of the drifted probability density function $F^{(t)}(x_1,x_2)$ and so it is valid to perform the convolution of this starting density with the diffusion density function $G(x_1,x_2,\Delta t)$, yielding the resulting density function $F^{(t+\Delta t)}$  \label{fig:2DMoateDiffusion Step}.}
	\end{center}
\end{figure}
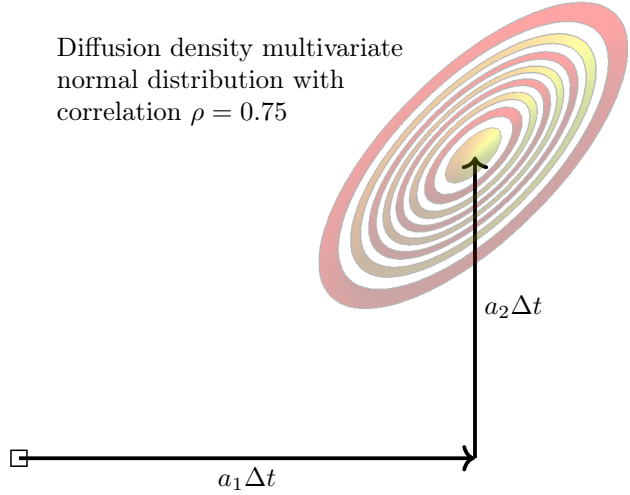

We also need to extend our notation from earlier so that the application of the drift terms in each dimension to the original distribution $F^t({\mathbf x})$, given by a vector ${\mathbf a}({\mathbf x}, t)$, is represented by the operator $\aplus$ so that 
\begin{equation}
	F^{(t+\Delta t)}({\mathbf x}) =  \left( F^{(t)}({\mathbf x}) \aplus {\mathbf a}({\mathbf x},t) \;\Delta t \right) \ast G({\mathbf x},\Delta t) 
\end{equation}
for the drift-first formulation of the multidimensional Moate Simulation.  Figure \ref{fig:2DMoateDiffusion Step} show a two dimensional example where a drift is performed $\left( F^{(t)}({\mathbf x}) \aplus {\mathbf a}({\mathbf x},t) \;\Delta t \right)$ and then a diffusion step ($G({\mathbf x},\Delta t)$) with a bivariate normal distribution that shows non-zero correlation.

\subsection{Local and Stochastic Volatility Models}
In local volatility models the diffusion term is a function of price $S_t$ and time $t$ where the only source of randomness comes from the Wiener increment of the price process.   Stochastic volatility models introduce additional sources of risk to model the time evolution of the volatility and this volatility is used in the resulting price (or interest rate, etc.) process.  Examples of stochastic volatility models include the  SABR Model\cite{HaganSABR}  and the Heston Model\cite{HestonModel1993}.

Consider the Heston Model with the form
\begin{eqnarray}
	dS_t &=& \mu S_t \; dt \; + \sqrt{\nu_t}  S_t \; dW_t^S \\
	d\nu_t &=&  \kappa(\theta -\nu_t)\; dt \; + \; \xi \sqrt{\nu_t} \; dW_t^\nu \label{eqn:HestonVariance}\\ 
	dW_t^S\; dW_t^\nu &=& \rho \; dt
\end{eqnarray}
If  volatility $\sigma$  is set to be the square root of the variance $\nu$  then use of It\^o's Lemma for each equation gives the system of equations 
\begin{eqnarray}
	\left( \begin{array}{c}
	d(\log S_t)  \\
	d\sigma_t 
\end{array} \right)
&=&
	\left( \begin{array}{c}
	 \mu - \frac{\sigma_t^2}{2}  \\
\frac{\kappa (\theta - \sigma_t^2)}{2 \sigma_t}    -\frac{\xi^2}{8\sigma_t}  \end{array} \right)
\; dt \; + \;
	\left( \begin{array}{cc}
	\sigma_t  &  0\\
	0         &  \frac{\xi}{2}  \end{array} \right) \; \cdot \;
	\left( \begin{array}{c}
	dW_t^S\\
	\;dW_t^\nu   \end{array} \right)
\\
	\left( \begin{array}{c}
		d(\log S_t)  \\
		d\sigma_t 
	\end{array} \right)
&=&
	\left( \begin{array}{c}
		\mu - \frac{\sigma_t^2}{2}  \\
		\frac{\kappa (\theta - \sigma_t^2)}{2 \sigma_t}    -\frac{\xi^2}{8\sigma_t}  \end{array} \right)
	\; dt \; + \;
	\left( \begin{array}{cc}
		\sigma_t  &  0\\
		0         &  \frac{\xi}{2}  \end{array} \right) \; \cdot \;
	\left( \begin{array}{cc}
	     1    &  0\\
	\rho      &  \sqrt{1-\rho^2}\end{array} \right) \; \cdot \;
	\left( \begin{array}{c}
		dW_t^{(1)}\\
		\;dW_t^{(2)}   \end{array} \right)
\end{eqnarray}
where we have used a Cholesky decomposition of the correlation matrix to transform the $\rho$ correlated Wiener increments $dW_t^S$ and $dW_t^\nu$ into independent Wiener increments $dW_t^{(1)}$ and $dW_t^{(2)}$.

The log price process depends upon the distribution of the volatility at time $t$, and the volatility process is a function of the other parameters alone (but independent of the price).  The volatility distribution can be simulated from time zero to any time $t$   so that the $\sigma_t$ distribution is available for the log price process.  Instead of treating $\sigma_t$ as a draw from that distribution, which is denoted here as $V^{(t)}(\sigma)$, we can consider the log price increment in the discrete time process, $\Delta (\log S_t)$, as a set of increments on its distribution at time $t$ each with a different volatility weighted by its probability,
\begin{equation}
	f^{(t+\Delta t)}(\log S)  =  
	\int_0^\infty \left( 
	    \left( f^{(t)}(\log S) \aplus  	\left(\mu - \frac{\sigma_{t}^2}{2}\right)\Delta t\right) \ast \sigma_{t} {\cal N}(0,  \Delta t) \right)   V^{(t)}(\sigma) \; d\sigma.
\end{equation}
Since the probability density function for volatility is already present on its own grid with associated probability density mass this can be approximated as the sum over the grid points $i \in n$ for the volatility probability density function as
\begin{equation}
	f^{(t+\Delta t)}(\log S)  =  
	\sum_i^n \left( 
	\left( f^{(t)}(\log S) \aplus  	\left(\mu - \frac{\sigma_{i,t}^2}{2}\right)\Delta t\right) \ast \sigma_{i,t} {\cal N}(0,  \Delta t) \right) V^{(t)}(\sigma_i) \; \Delta\sigma.
\end{equation}
The summation, and in the limit the integration too, can be considered as a probability weighted sum over a set of states of the world with different volatility.
This numeric approach is tractable and each volatility probability weighed sum of distributions arising from different volatility values and in practical algorithmic terms this is highly parallelisable.

Consider also the SABR Model\cite{HaganSABR} 
\begin{eqnarray}
	dF_t &=& \sigma_t (F_t)^\beta \; dW_t^{(F)} \label{eqn:SABRF}\\
	d\sigma_t &=& \alpha \sigma_t \; dZ_t^{(\sigma)} \label{eqn:SABRsigma}\\ 
	dW_t^F\; dW_t^\sigma &=& \rho \; dt
\end{eqnarray}
The forward process $F_t$ depends on the stochastic volatility process $\sigma_t$, with $\alpha,\beta,\rho$ all being constant parameters within the model. For this demonstration we assume $\rho=0$.  The $\sigma_t$ process, equation \ref{eqn:SABRsigma}, can be transformed into a $\log \sigma_t$ process using It\^o's Lemma,
\begin{equation}
	d(\log \sigma_t) = -\frac{\alpha^2}{2}\; dt \;+ \; \alpha\;  dZ_t^{(\sigma)}.
\end{equation}   
A change in variable $q(F) =  F^{(1 - \beta)}/(1 - \beta)$  for $\beta \ne 0$, or $q(F) =\log(F)$  for $\beta=0$, gives a process with identical diffusion functional coefficient for any value of $q$ so that applying It\^o's lemma to equation \ref{eqn:SABRF} yields
\begin{equation}
	dq_t =
	\begin{cases}
		-\frac{1}{2} \frac{\beta}{1-\beta}  \frac{\sigma_t^2}{q_t} \;dt\;  +\; \sigma_t  \; dW_t^{(F)} ,& \text{if } \beta \ne 0 \\
	    -\frac{\sigma_t^2}{2}\;dt \; + \sigma_t dW_t^{(F)},              & \text{if } \beta = 0 .
	\end{cases}
	%
	%dq_t &= -\frac{1}{2} \frac{\beta}{1-\beta}  \frac{\sigma_t^2}{q_t} \;dt\;  +\; \sigma_t  \; dW_t^{(F)}  & \hbox{for\ }  \beta \ne 0 \\
	%dq_t &= -\frac{\sigma_t^2}{2}\;dt \; + \sigma_t dW_t^{(F)}  & \hbox{for\ }  \beta = 0 .
\end{equation}
The log volatility could be simulated but it is also available in an analytic form thus we know at $t$ what the distribution of $\sigma_t$ is exactly.  As in the case for the Heston model shown above,  where here $V_t(\sigma)$ is the probability distribution derived from the distribution of $\log \sigma$ at time $t$ by change of variable, we obtain a single Moate Simulation step given by 
\begin{equation}
	f^{(t+\Delta t)}(q) =
	\begin{cases}	
		\int_0^\infty \left( 
		\left( f^{(t)}(q) \aplus  	\left(-\frac{1}{2} \frac{\beta}{1-\beta}  \frac{\sigma_t^2}{q_t}\right)\Delta t\right) \ast \sigma_{t} {\cal N}(0,  \Delta t) \right)   V^{(t)}(\sigma_t) \; d\sigma_t,& \text{if } \beta \ne 0 \\
				\int_0^\infty \left( 
		\left( f^{(t)}(q) \aplus  	\left(-\frac{\sigma_t^2}{2}\right)\Delta t\right) \ast \sigma_{t} {\cal N}(0,  \Delta t) \right)   V^{(t)}(\sigma_t) \; d\sigma_t,& \text{if } \beta = 0 .
	\end{cases}
\end{equation}
These integrations can be approximated as before where $\sigma$ is represented on a grid with points $i \in n$
\begin{equation}
	f^{(t+\Delta t)}(q) =
	\begin{cases}	
		\sum_i^n \left( 
		\left( f^{(t)}(q) \aplus  	\left(-\frac{1}{2} \frac{\beta}{1-\beta}  \frac{\sigma_t^2}{q_t}\right)\Delta t\right) \ast \sigma_{ti} {\cal N}(0,  \Delta t) \right)   V^{(t)}(\sigma_i) \; \Delta\sigma,& \text{if } \beta \ne 0 \\
		\sum_i^n \left( 
		\left( f^{(t)}(q) \aplus  	\left(-\frac{\sigma_t^2}{2}\right)\Delta t\right) \ast \sigma_{ti} {\cal N}(0,  \Delta t) \right)   V^{(t)}(\sigma_i) \; \Delta\sigma,& \text{if } \beta = 0 .
	\end{cases}
\end{equation}
which again is summing the probability densities over the $n$ states, each representing a state with probability $V_t(\sigma_i) \,\Delta \sigma$ with a different volatility $\sigma_i$.  With a probability density function known at $t + \Delta t$ the next time set in the simulation can be taken and this is repeated until the desired time horizon. 

This section has sketched an approach that uses the Moate Simulation for two stochastic volatility models;  this approach can be extended to  similar local or stochastic volatility models where there is no recourse to analytic solutions.

\section{Conclusions}
The key insight provided by the Moate Simulation approach introduced here is that convolution techniques can be applied to a distribution function where there are time or spatial dependencies in the drift and diffusion coefficients for stochastic processes. One strength of this approach is that highly accurate numerical results can be obtained. This allows, in the case of finance, the discounting of the evolved terminal spatial distribution to provide an accurate present value of the financial instrument or derivative under consideration.  The approach is general and can be applied to processes which describe prices, or other modelled features like rates or volatility. Thus, the approach can provide new ways of eliciting accurate results where closed-form analytic solutions have not been discovered. 

Additionally, the numerical advantages which Moate Simulation provides compared to Monte Carlo techniques  vis-a-vis estimation error will be welcomed in many areas of simulation; in finance these include derivative pricing, obtaining estimates of capital requirements to cover counterparty credit risk in institutional portfolios, in financial models for stress testing, Value-at-Risk and Expected Shortfall estimation, or for economic forecasting.  In particular the ability of Moate Simulation to represent more accurate tail distributions than Monte Carlo Simulation is notable. Practically, the simulation of stochastic processes can make use of modern hardware like Graphic Processing Units where  libraries have been parallelised and optimised for Fourier transformations and convolution.  

Moate Simulation will allow researchers more freedom to select the stochastic models they want to use on the basis of suitability rather than restricting their study to models which may have known closed-form or approximate analytic solutions but which may be limited by their inherent assumptions.

\newpage

 % Entries are in the references.bib file

\end{document}